\documentclass[annual]{acmsiggraph}

\title{Medial Meshes for Volume Approximation}

\author{
{Feng Sun\hspace{.5in}
Yi-King Choi\hspace{.5in}
Yizhou Yu\hspace{.5in}
Wenping Wang}\thanks{e-mail:[fsun,ykchoi,yzyu,wenping]@cs.hku.hk}\\
The University of Hong Kong
}

\pdfauthor{Feng Sun,Yi-King Choi,Yizhou Yu,Wenping Wang}


\usepackage{url}

\usepackage{breakurl}

\usepackage{amsmath}
\usepackage{amssymb}
\usepackage{parskip}
\usepackage{bm}
\usepackage{picins}
\usepackage{color}
\usepackage{enumerate}
\usepackage{bm}
\usepackage{subfigure}
\usepackage{enumitem}
\usepackage{multirow}
\usepackage{rotating}
\usepackage{array}
\usepackage{afterpage}
\usepackage{algorithmic}
\usepackage{algorithm}

\def \mp{\mathbf{p}}
\def \mq{\mathbf{q}}

\def \mS{\mathbf{S}}

\def \mC{\mathbf{C}}

\def \mv{\mathbf{v}}

\def \cG{\mathcal{G}}
\def \cN{\mathcal{N}}
\def \cV{\mathcal{V}}

\definecolor{light-gray}{gray}{0.8}

\begin{document}

\teaser{
\captionsetup{type=figure}
\includegraphics[width=\linewidth]{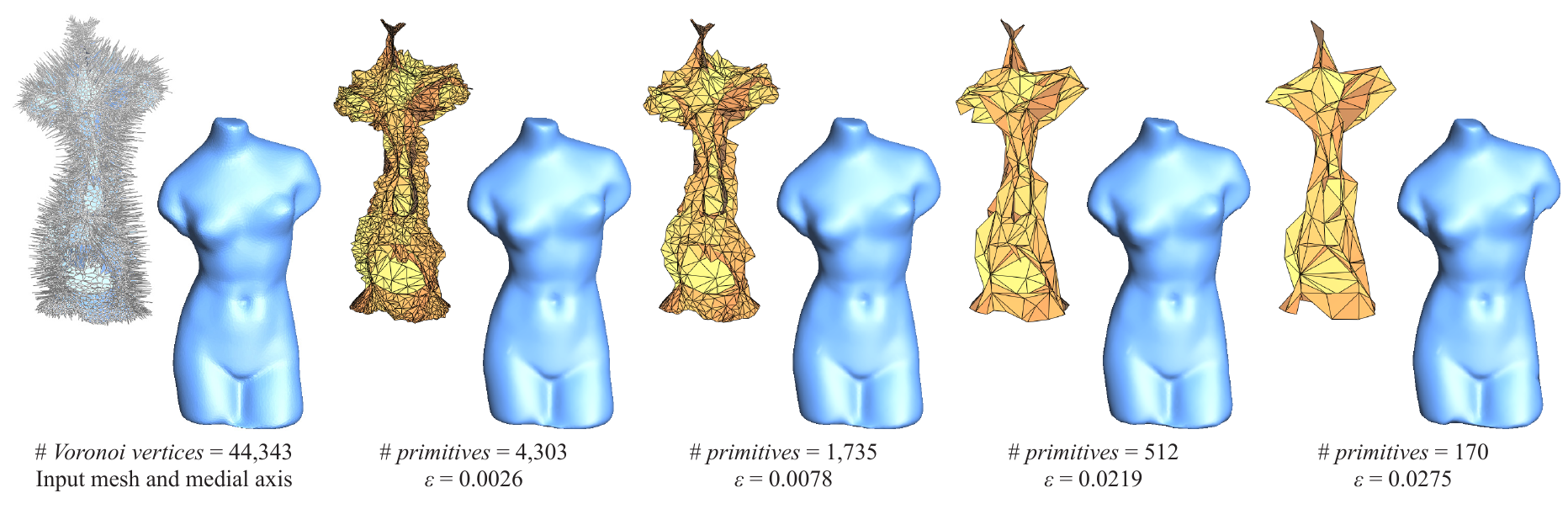}
\caption{A series of volume approximations based on progressively simplified medial meshes computed by our method. The approximation error is measured by the one-sided Hausdorff distance from the original shape to approximate volumes. }\label{fig:teaser}
}

\maketitle

\begin{abstract}
Volume approximation is an important problem found in many applications of computer graphics, vision, and image processing. The problem is about computing an accurate and compact approximate representation of 3D volumes using some simple primitives. In this study, we propose a new volume representation, called {\em medial meshes}, and present an efficient method for its computation. Specifically, we use the union of a novel type of simple volume primitives, which are spheres and the convex hulls of two or three spheres, to approximate a given 3D shape. We compute such a volume approximation based on a new method for medial axis simplification guided by Hausdorff errors. We further demonstrate the superior efficiency and accuracy of our method over existing methods for medial axis simplification.

\end{abstract}


\keywordlist



\section{Introduction}
\label{sec:Introduction}


Volume representation for 3D shapes is ubiquitous in computer graphics, as well as other fields of science and engineering, such as CAD/CAM, vision and image processing. Three important considerations in the study of volume representation are simplicity, accuracy, and efficiency. Specifically, with a specified type of volume representation, we wish to have the following properties: (1) {\bf simplicity}: simple primitives are used to approximate an arbitrary 3D shape in order to facilitate subsequent processing, such as rendering and simulation; (2) {\bf accuracy and efficiency}: it is necessary to have an {\em efficient} method for computing an {\em accurate} approximation of an arbitrary 3D shape using the specified representation.

We propose a new approach to volume representation in this paper. We first present a new class of simple volume primitives and propose to use their union to approximate any given 3D shape. These primitives include spheres, and convex hulls of two or three spheres (see Figure~\ref{fig:one triangle}). These primitives are simple to analyze and process since their boundary surfaces are defined by spherical caps, conical patches, and triangle faces, and therefore allow fast geometric processing.

Note that these primitives are naturally related to the {\em medial axis transform} (MAT) of 3D shapes. Briefly speaking, the exact MAT of a 3D object is a set of infinitely many {\em maximal spheres} (also called {\em medial spheres}) whose union is the given 3D object. In computational practice, the MAT is often approximated by the union of finitely many spheres. Our insight is that the MAT can be approximated efficiently using a 2D non-manifold simplicial complex, to be called a {\em medial mesh}, which consists of line segments and triangle faces. The vertices of a medial mesh are sampled medial spheres of the MAT, and the linear interpolation of these vertices over line segments or triangles of the medial mesh defines the convex hulls of two or three medial spheres. Because of this connection between the medial mesh and MAT, in the following we shall mainly be concerned with the simplification of the medial axis of a 3D object for computing medial meshes. 

To convert the above observations into an effective computation method, we need to address the following issues: (1) How to simplify a densely sampled MAT of a given 3D object to produce a compact representation with a reduced data size? (2) How to ensure that the resulting approximation results in an accurate representation of the given 3D object? In addition, we need to address the notorious instability issue in medial axis computation in the process.

\begin{figure}[htbp]
\hspace*{.05\linewidth}
\begin{minipage}{.35\linewidth}
\centering
	\includegraphics[width=\linewidth]{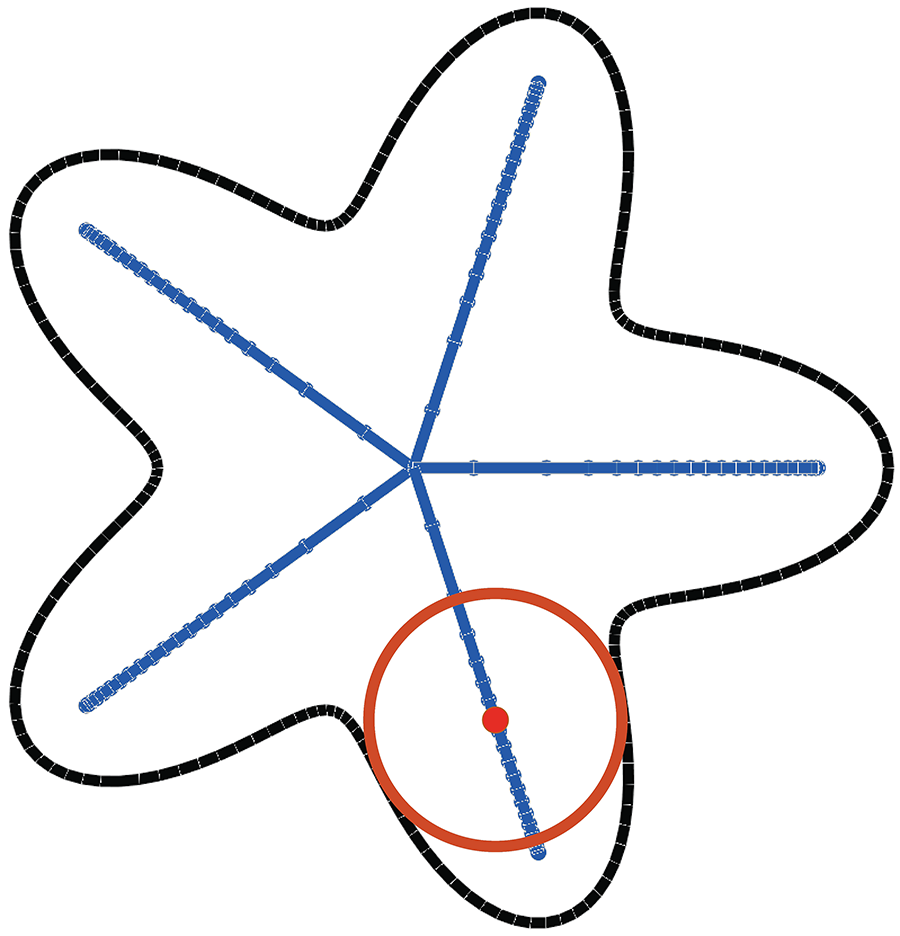}\\
	({\em a})
\end{minipage}
  \hfill
\begin{minipage}{.35\linewidth}
\centering
  \includegraphics[width=\linewidth]{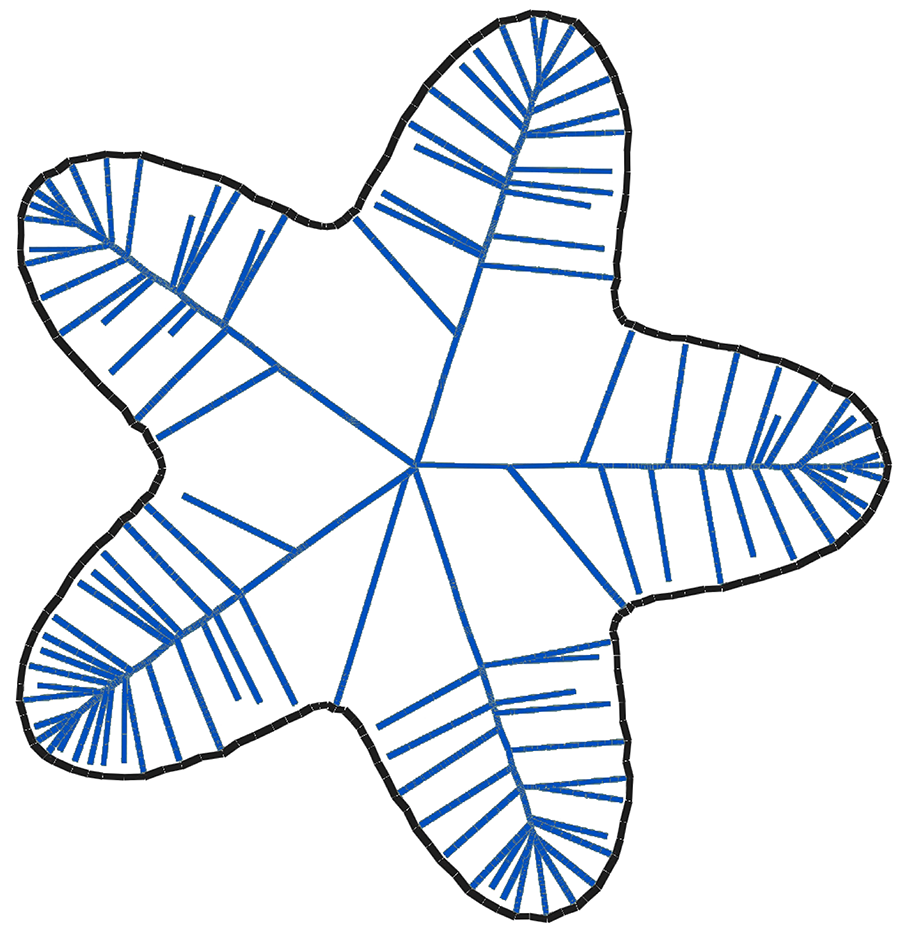}\\
  ({\em b})
\end{minipage}
\hspace*{.05\linewidth}
\caption{{\bf The instability of medial axes}. (a) The medial axis of a flower shape with a smooth boundary. (b) The medial axis of the same shape with a noisy boundary. }
\label{fig:unstable medial axis illustration}
  \end{figure}

The medial axis of a solid object in $\mathbb{R}^d$, $d=2$ or $3$, is the set of points having at least two closest points on the object's boundary, in other words, it comprises the center of the spheres (or circles in 2D) which are contained in the object and touch the object's boundary at two or more points. These spheres are called the {\em medial spheres} or ({\em medial circles} in 2D). See \autoref{fig:unstable medial axis illustration}(a) for a 2D illustration. The medial axis transform (MAT) consists of two parts, a medial axis and a radius function, which encodes the radii of the associated medial spheres. As an intrinsic shape representation, the MAT has proven extremely useful for shape analysis and synthesis tasks, such  as the approximation, description, recognition and retrieval of shapes as well as topology representation and data reduction of complex models. In the following we shall study the computation of the MAT in order to generate a simple and compact volume representation.

Despite its potential utility, the application of the MAT has been hindered by its instability and redundancy. First, the MAT is notoriously sensitive to noisy, small perturbations on the shape boundary. As shown in \autoref{fig:unstable medial axis illustration}, a slightly noisy shape boundary leads to a medial axis with numerous undesired branches. Such instability plagues the application of the MAT on real-world data with noise. Second, the MAT is in general a continuum comprising infinitely many points with a simple analytical expression. As a result, in practice a medial axis is commonly approximated with a dense set of sample points, which amounts to using the union of a large number of medial spheres to reconstruct the original shape. This leads to inefficiency and poor accuracy when computing with the MAT.

We shall use a triangular mesh to approximate the MAT, and call it a {\em medial mesh}. Unlike previous methods that use a discrete set of individual spheres, the medial mesh uses linear interpolation of sampled medial spheres to approximate the MAT, resulting in a more accurate and compact approximation.

We shall present an effective algorithm for computing a stable and compact medial mesh. Our algorithm not only cleans up the topology of the medial axis by pruning unstable branches, but also produces a compact representation by reducing the number of sample points on the medial axis while ensuring shape approximation accuracy of the medial mesh by observing a specified error bound during simplification. Analogous to mesh simplification, our algorithm progressively simplifies an initial medial mesh of the input shape by iteratively contracting selected edges until the approximation error reaches a predefined threshold. Because of the use of sphere interpolation, this algorithm is capable of drastically reducing the number of vertices and edges in the medial mesh while faithfully representing the original shape. Experiments and comparisons indicate that our simplified medial mesh can achieve the same shape approximation error with orders-of-magnitude fewer primitives than by existing medial axis pruning techniques. This leads to an accurate and compact volume presentation using the proposed simple primitives.

We shall demonstrate the relevance of medial meshes in two shape modeling tasks. In shape approximation, the mesh is able to achieve much smaller approximation errors with the same number of primitives in comparison to state-of-the-art methods based on spherical approximation. In free-form shape deformation, the medial mesh can serve as the compact embedded structure of an object to offer flexibility in bending and stretching while faithfully preserving the object thickness than state-of-the-art deformation techniques.

\section{Related Work}

\paragraph{Volume approximation}

Shape or volume approximation with simple primitives is a fundamental problem in computer graphics and computational geometry, as it enables simplified and efficient computations in various applications. Focusing on fast collision detection, Hubbard~\shortcite{Hubbard1996} proposes the {\em sphere tree} as a hierarchical structure which is built based on mid-axis surface computation.  Bradshaw and O'Sullivan \shortcite{Bradshaw2004} extend this method using adaptive medial axis approximation and improves the bounding efficiency by with an iterative greedy approach.
Wang et al.~\shortcite{Wang2006} adopt a variational approach that minimizes the outsize volume of the spheres to obtain a set of sphere whose union bounds and approximates a given object, which is used for fast shadow computation~\cite{Ren2006}. Based on the medial axis, Stolpner et al. \shortcite{Stolpner2011a} present a method for sampling internal medial spheres to obtain a tight approximation of a given shape. All these methods use the union of spheres to approximate a given object and therefore often need a large number of spheres to achieve satisfactory approximation. In contrast, we propose to use simple primitives that are linear interpolations of spheres to define a volume approximation. 

Apart from spheres, ellipsoids are also effective primitives for volume approximation.  Bischoff and Kobbelt~\shortcite{BischoffK2002} use ellipsoids to cover a 3D object.  Specifically designed for surface reconstruction in robust geometry transmission, the method yields a decomposition that contains a large number of ellipsoids than necessary for tight bounding. Lu et al.~\shortcite{LuCWK2007} propose a variational approach to computing an optimal segmentation of a 3D shape for computing a bounding volume defined by a union of ellipsoids.  With an error-driven segmentation refinement, the method is capable of achieving shape approximations that satisfy a user-specified volumetric error.

\paragraph{Medial axis simplification}
The medial axis transform was first proposed by Blum~\shortcite{Blum1967} as a tool for biological applications. It has been proved that the MAT is injective, therefore a complete shape descriptor. The medial representation captures the shape intrinsically by encoding the local thickness and symmetry, and finds applications in shape matching, shape recognition, shape retrieval \cite{Kazhdan2003,Siddiqi2008a,Siddiqi2008b}, to name a few. However, the medial axis is inherently unstable, that is, a small perturbation to the shape boundary may introduce a large change of its medial axis. In graphics, models often have a boundary representation, such as triangle meshes. The exact medial axis of such models \cite{Culver2004,Aichholzer2010} typically has many undesired branches, therefore are not suitable for further applications. To overcome the instability of the medial axis, several methods have been proposed to remove the unstable spikes, a process often referred to as medial axis pruning.

\textit{Angle-based filtering.} \quad%
For every point on a medial axis, angle-based methods \cite{Attali1996,Amenta2001b,Dey2002,Foskey2003} compute the angle spanned by its closest points on the shape boundary. All the points on the medial axis with a spanned angle less than a user-specified threshold are removed. The points surviving this removal make up the filtered medial axis. Although the angle criterion can preserve local features, angle-based filtering often yields a simplified medial axis with a different topology from the input one~\cite{Miklos2010}.


\textit{The $\lambda$ medial axis.} \quad%
Another criterion for medial axis filtering is the circumradius of closest points of a medial point. Such a filtered medial axis is also known as the $\lambda$ medial axis \cite{Chazal2005,Chaussard2009}. All medial points with a circumradius smaller than a given threshold $\lambda$ are removed. It has been proven that such filtration preserves the topology for small $\lambda$ \cite{Chazal2005}. However, this circumradius criterion does not work well on shapes with features at different scales. Small values of $\lambda$ cannot remove noise near large-scale features while increasing $\lambda$ would eliminate small-scale features. As a result, there may exist large discrepancies between the original shape boundary and the shape boundary reconstructed from the simplified medial axis.

\textit{Scale axis transformation.} \quad%
Miklos et al.\ \shortcite{Miklos2010} proposes a method based on the scale axis transformation (SAT) \cite{Giesen2009}. For a given medial axis, all medial spheres are first scaled by a factor $s$ larger than $1$. A scaled medial sphere is removed if it is contained in another scaled medial sphere. Then, the union of all remaining medial spheres generates a new shape, whose medial axis is further subject to topology-preserving angle filtering. The final result is obtained by shrinking all medial spheres of the simplified medial axis by the factor ${1}/{s}$. This SAT based method often generates results better than earlier techniques. However, it does not preserve the shape topology as it may fill in narrow gaps or small holes by the first dilating step.


\textit{Feature-based simplification.} \quad%
Another paradigm of medial axis simplification is taken by Tam and Heidrich~\shortcite{TamH2003} to achieve high-level feature-based shape simplification.  The medial axis is decomposed into manifold sheets as parts, each of which corresponds to a feature of the input shape.
The parts are then pruned based on significant measures using triangle count and the volume of each part.  While the method can remove insignificant shape features as well as components that are smaller than a volume threshold, it does not reduce the number of triangles on each of the remaining manifold sheet.  In other words, the geometric complexity of these parts remain intact.  The resulting feature filtered shape can be one with a medial axis still having lots of triangles.
Our method, on the other hand, outputs a simplified medial axis by reducing the number of geometries used in the medial mesh but at the same time attaining an approximation error up to a user defined threshold. The output is therefore an error bound medial shape representation with a medial axis with much fewer geometries.

\textit{Interpolation of medial spheres.} \quad%
A common scheme shared by all previous pruning methods is that they discretize a medial axis into a set of points only, that is, they approximate the input shape using a discrete set of individual spheres. Such a discretization simplifies algorithm design but gives rise to artifacts or large shape approximation errors. In contrast, with our simplified medial mesh we use linear interpolations of spheres to approximate the input shape to achieve both high visual and numerical fidelity. Note that our medial mesh simplification algorithm is quantitatively controlled by a shape approximation error metric.

\textit{M-rep.} \quad%
It is a little surprising that there has been no previous study in literature on how to compute a compact and stable medial axis representation based on the interpolation of medial sphere. A related work is M-rep by Pizer et al.\ ~\shortcite{pizer1999segmentation}, which proposes the compact spline approximation to the medial axis of 3D objects for shape analysis in medial imaging. However, M-rep considers only the special case that the 3D object is simple enough to allow its medial axis to be approximated by a single patch of tensor product B-spline surface. Furthermore, it assumes that this tensor product B-spline surface patches is manually specified and does not consider how to extract such a spline representation automatically. The idea of medial mesh is inspired by the M-rep, but our goal to is compute an accurate and compact medial axis approximation of {\em any 3D object} based piecewise linear interpolation of medial spheres.

\section{Definition of Medial Meshes}

The MAT of a 3D shape is in general a 2D non-manifold surface embedded in 4D space, since each point of the medial axis consists of the coordinates of its 3D position and the radius of its associated medial sphere. As demanded by many applications in shape analysis and synthesis, we are concerned with accurate and compact representation of this medial axis surface. For this purpose we propose to study the piecewise linear approximation to the medial axis surface.

\begin{figure}
\centering
\begin{minipage}{.32\linewidth}
\centering
\includegraphics[width=\linewidth]{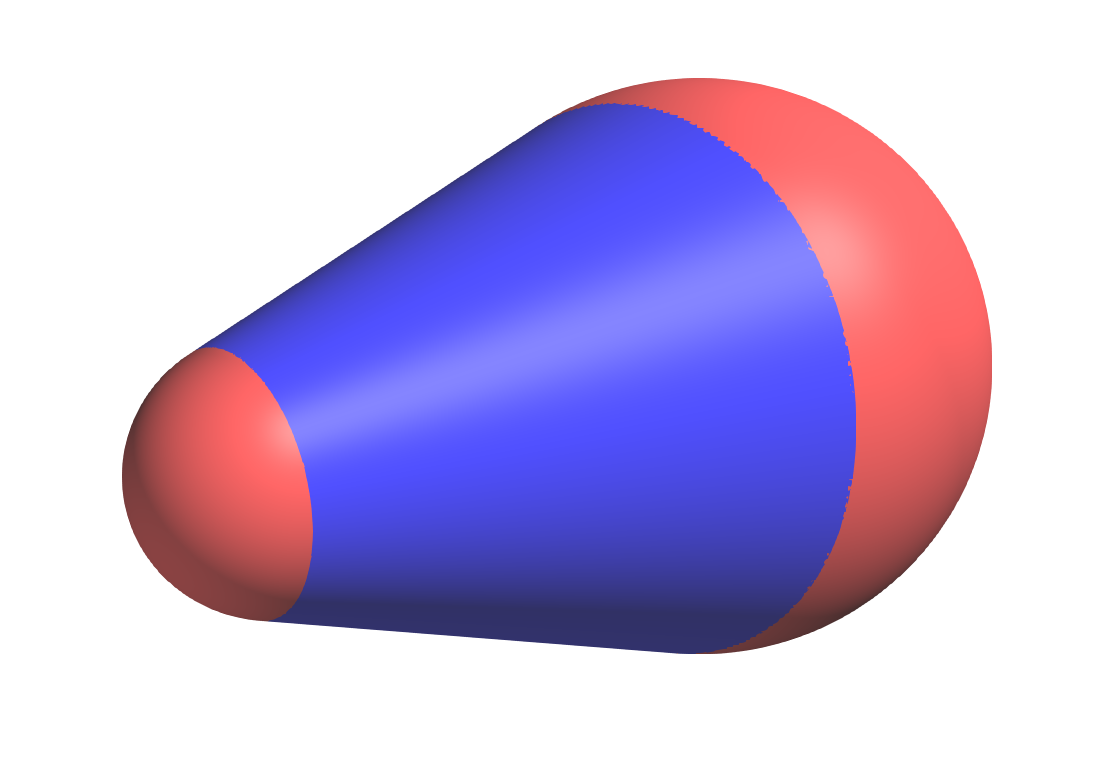}\\
{\small ($a$)}
\end{minipage}\hspace{5ex}
\begin{minipage}{.32\linewidth}
\centering
\includegraphics[width=\linewidth]{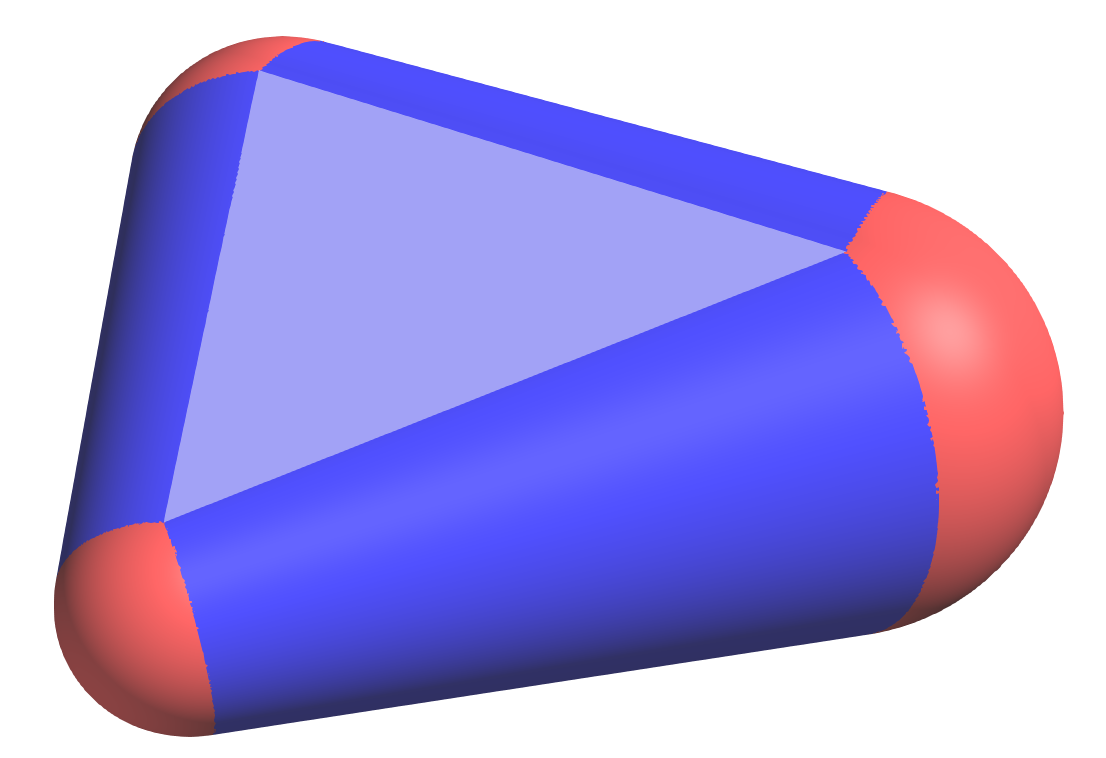}\\
{\small ($b$)}
\end{minipage}
\caption{($a$) The enveloping primitives of two spheres represented by an edge of the medial mesh. ($b$) The enveloping primitives of three spheres represented by a triangle face of the medial mesh.}
\label{fig:one triangle}
\end{figure}

The {\em medial mesh} of a 3D object is a 2D simplicial complex approximating the medial axis of the object. A vertex of a medial mesh is called a {\em medial vertex} and is a 4D point $\mv = (\mp, r)$, where $\mp$ is the 3D position of the vertex and $r$ its associated radius value.  The volume primitive of a medial vertex is the medial sphere with center $\mp$ and radius $r$. An edge $e = \{\mv_1, \mv_2\}$ of the medial mesh is called an {\em medial edge}, represented by $(1-t)\mv_1 + t\mv_2, t\in[0,1]$, which is a convex interpolation of its end points $\mv_1$ and $\mv_2$. Geometrically,  the volume primitive generated by a medial edge is the convex hull of the two medial spheres defined by $\mv_1$ and $\mv_2$, or a swept volume generated by interpolating the two spheres (see Fig.~\ref{fig:one triangle}(a)). The most general type of elements of a medial mesh is a triangle face, called the {\em medial face}. It is defined by  $f=\{\mv_1, \mv_2, \mv_3\}$, which is the convex combination of the three medial points, i.e., $a_1\mv_1 + a_2\mv_2 + a_3\mv_3$, where the $a_i$ are the barycentric coordinates satisfying $a_1+a_2+a_3=1$ and $a_i \ge 0 \,(i = 1,2,3)$. Geometrically, the volume primitive generated by a medial face is the convex hull of the three medial spheres defined by the vertices $\mv_1, \mv_2, \mv_3$ (see Fig.~\ref{fig:one triangle}(b)).

\begin{figure}[htbp]
	\begin{minipage}{.3\linewidth}
	\centering
  \includegraphics[width=\linewidth]{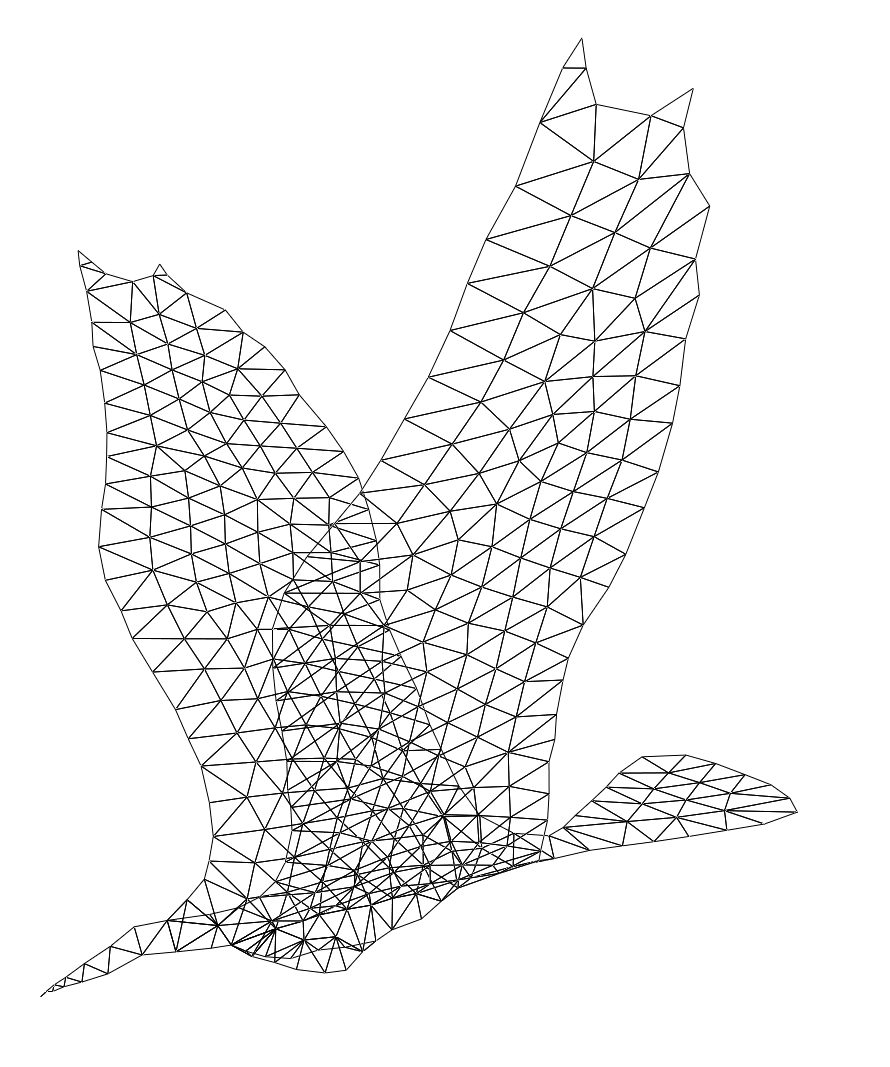}
  \end{minipage}
 \hfill
	\begin{minipage}{.3\linewidth}
	\centering
  \includegraphics[width=\linewidth]{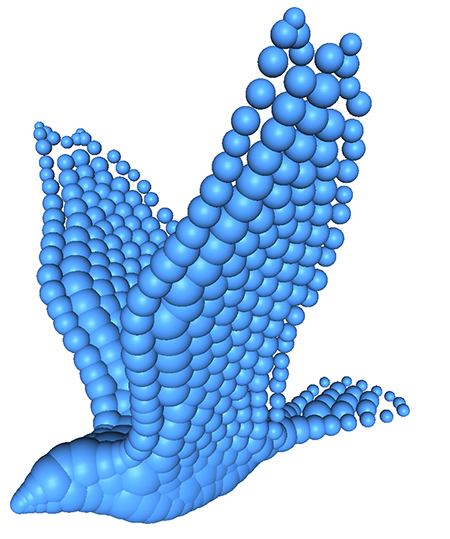}
  \end{minipage}
 \hfill
	\begin{minipage}{.3\linewidth}
	\centering
  \includegraphics[width=\linewidth]{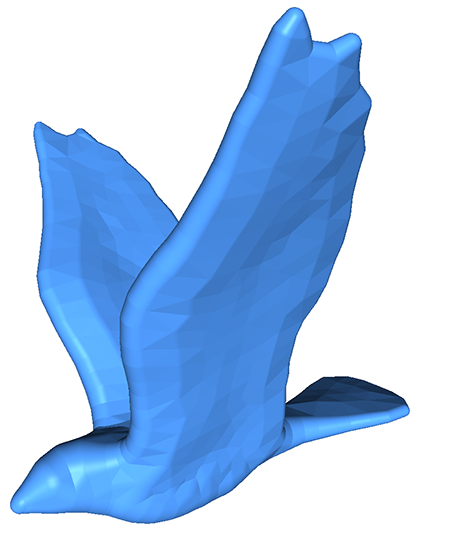}
  \end{minipage}
\caption{{\bf An illustrative example}. From left to right: A medial mesh of a bird shape with 400 medial vertices. Reconstruction by the union of the 400 sampled medial spheres. Reconstruction by the union of the enveloping primitives of the medial mesh.}
\label{fig:medial mesh}
\end{figure}

The volume primitives associated with the vertices, edges and faces of a medial mesh will be called {\em enveloping primitives}. The medial mesh then represents a 3D object that is the union of all its enveloping primitives. This representation is much more compact that those previous approaches of using the union of sampled individual medial spheres to approximate the medial axis, as illustrated in Fig.~\ref{fig:medial mesh}.

The analogue of the medial mesh for 2D shapes is a graph ($V, E$) consisting of a set $V$ of medial vertices and a set $E$ of medial edges connecting medial vertices. Here the interpolation of medial spheres for a medial mesh in 3D is replaced by interpolation of medial circles.

\section{Medial Mesh Computation}

We now discuss how to compute a concise and stable medial mesh to robustly represent and approximate a given shape. Given the boundary mesh of a 3D shape, using the mesh vertices or a set of sampled point on the mesh as input, we first compute an initial medial axis using the Voronoi-based method~\cite{Amenta1998,Attali1997}. This initial medial axis is a noisy and dense mesh representation of the medial axis and will serve as the initial medial mesh. Starting from it, we iteratively contract selected edges to progressively simplify the medial mesh until a user-specified approximation error has been reached. Our simplification strategy is driven by approximation error of the simplified medial mesh to the input shape. So it is capable of reducing the number of medial vertices as well as eliminating unstable spikes, which contribute negligibly to the shape boundary and therefore their removal gives rise to very limited shape approximation error.

\begin{figure*}
\begin{minipage}{.32\linewidth}
\centering
\includegraphics[width=\linewidth]{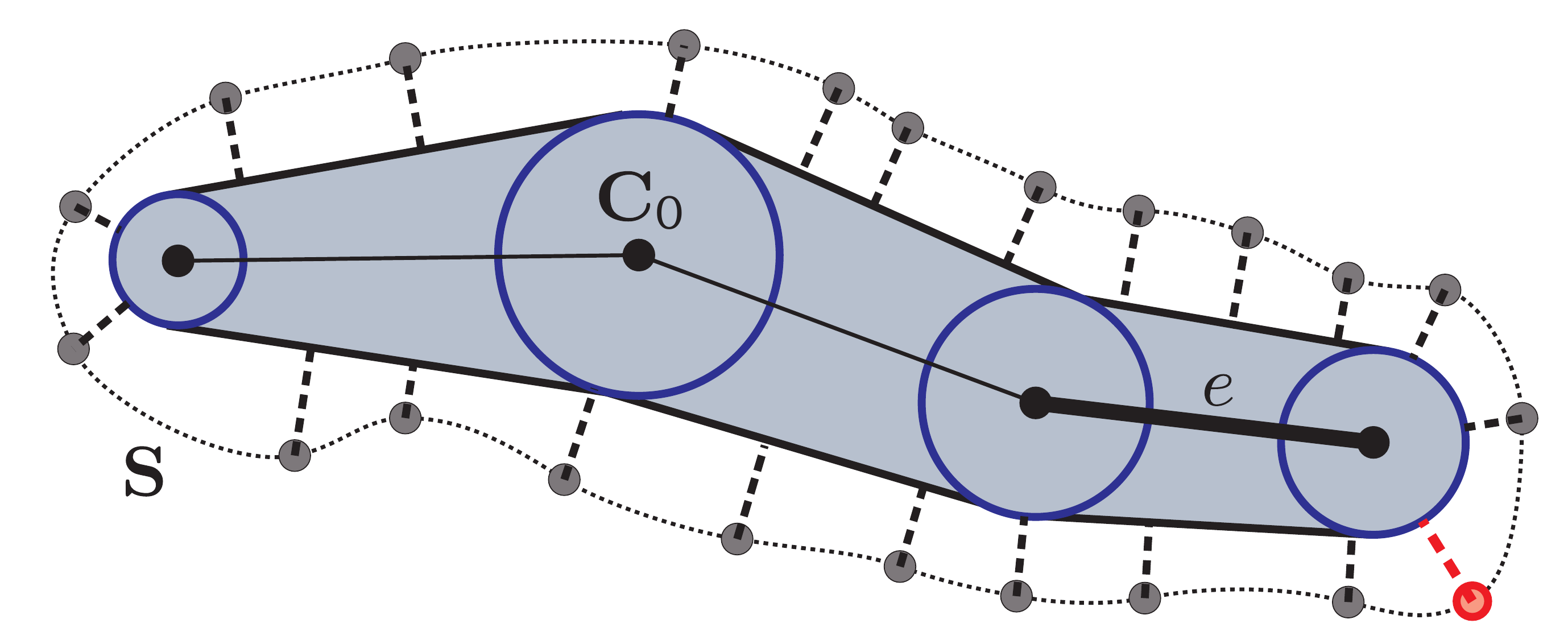}\\
($a$)
\end{minipage}
\hfill
\begin{minipage}{.32\linewidth}
\centering
\includegraphics[width=\linewidth]{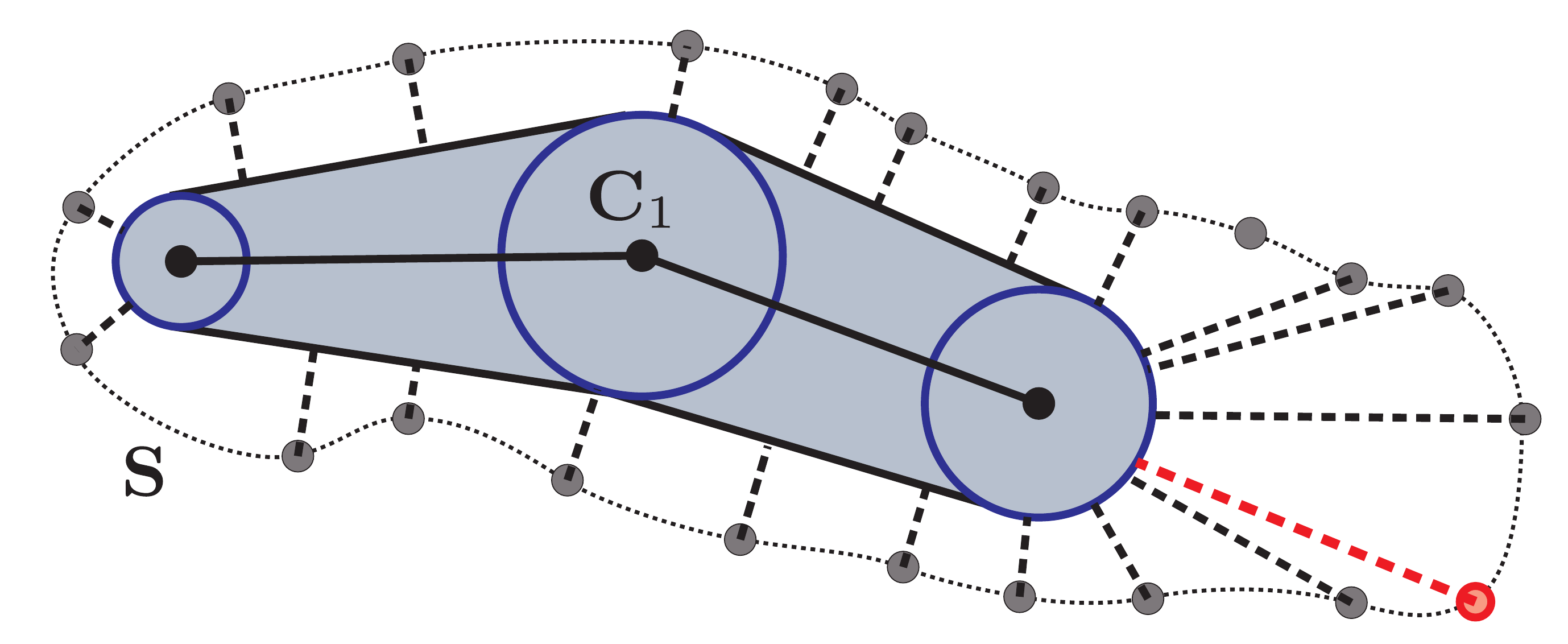}\\
($b$)
\end{minipage}
\hfill
\begin{minipage}{.32\linewidth}
\centering
\includegraphics[width=\linewidth]{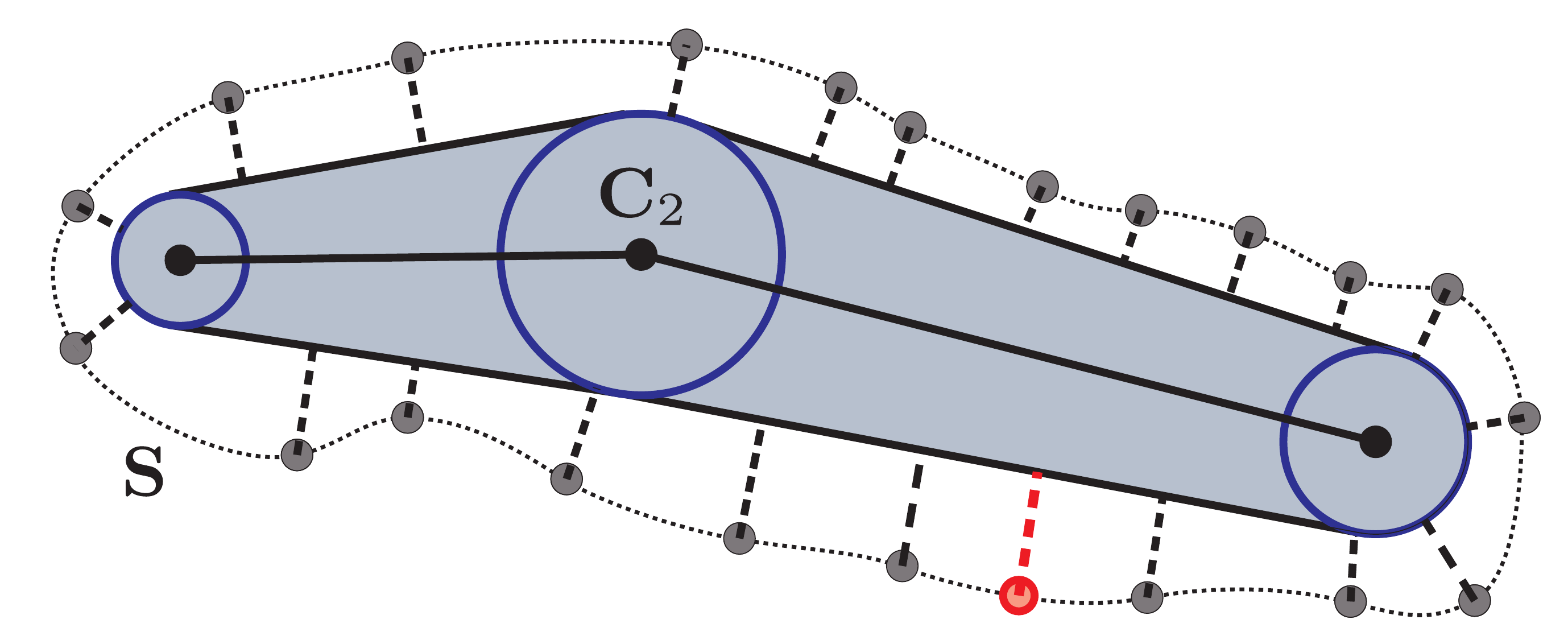}\\
($c$)
\end{minipage}
\caption{ {\bf Post-contraction error of an edge}. ($a$) The post-contraction error of an edge $e$ in the medial mesh $\mC_0$ approximating a shape $\mS$ is to be evaluated. The shape approximation error is highlighted in red. ($b$) Contract $e$ by merging the right vertex with the left one, resulting in a medial mesh $\mC_1$. ($c$) Contract $e$ by merging the left vertex with the right one, resulting in a medial mesh $\mC_2$.  The error associated with  $\mC_2$ in ($c$) is chosen as the post-contraction error of the edge $e$ since it is smaller than the error associated with $\mC_1$ in ($b$). }
\label{fig: 2d_envelope_contraction}
\end{figure*}

We now give the details of our simplification method. Given a 3D volumetric shape $S$, let $\mC_0$ denote a medial mesh approximating the medial axis of $S$. Let $\mS_0$ denote the union of all the enveloping primitives of the medial mesh $\mC_0$.  The boundary surface of $\mS_0$, denoted $\partial \mS_0$, is the surface reconstructed from the medial mesh $\mC_0$ to approximate the boundary $\partial \mS$ of the given shape $\mS$. The approximation error $\epsilon$ of the medial mesh $\mC_0$ with respect to $\mS$ is measured by the Hausdorff distance between $\partial \mS_0$ and the original boundary surface $\partial \mS$. Suppose that $\partial \mS$ is sampled by a dense set of points $\{\mq_i\}$. Then the approximation error $\epsilon$ is given by $\max_i\{d(\mq_i,\partial \mS_0)\}$ (\autoref{fig: 2d_envelope_contraction}(a)), where $d(\mq_i,\partial \mS_0)$ is the distance from the point $\mq_i$ to the reconstructed surface $\partial \mS_0$, and can be computed by taking the
minimum of the signed distances from $\mq_i$ to the spheres, truncated cones and triangles composing the boundary surfaces of the individual enveloping primitives.



Given a medial mesh $\mC_0$ whose envelope $\mS_0$ approximates the input shape $\mS$, contracting a medial edge $e \in \mC_0$ gives rise to a new medial mesh $\mC_1$. In general, $\mC_1$ has a larger approximation than $\mC_0$ and there is only a local region $R$ on the original boundary surface $\partial \mS$ near the contracted edge $e$  that contributes to this increased error. The mesh vertices in $R$ are those whose closest points on $\partial \mS_0$ lie on an enveloping primitive of $\mC_0$ that involves the edge $e$.


We define the {\em post-contraction error} of $e$ as the local shape approximation error caused by the contraction of $e$, i.e. $E_{contr}(e) = \max\{d(\mp, \partial \mS_1)| \mp \in R\}$, measuring the approximation error of the region $R$ on $\partial \mS$ by the corresponding region on $\partial \mS_1$ which is the boundary surface of the update medial mesh $\mC_1$, as shown in the 2D illustration in~\autoref{fig: 2d_envelope_contraction}. In \autoref{fig: 2d_envelope_contraction}(a), $\mC_0$ consists of three edges, and the edge whose post-contraction error needs to be evaluated is rendered with a thick line, denoted $e$. There are two ways to contract $e$, namely merging the left vertex with the right one, and vice versa. These two possible ways of contracting $e$ and their associated errors are shown in Figures~\ref{fig: 2d_envelope_contraction}(b) and (c), respectively. Since merging the left vertex to the right vertex leads to a smaller error associated, this error indicated in Figure~\ref{fig: 2d_envelope_contraction}(c) is chosen as the post-contraction error of the edge $e$.

Our simplification algorithm first computes the post-contraction error for every edge in the initial medial mesh $\mC_0$. It then iteratively contracts the edge with the smallest post-contraction error.
When an edge $(\mv_i,\mv_j)$ is contracted, the two medial vertices merges and all edges incident to $\mv_i$ or $\mv_j$, as well as their associated enveloping primitives (i.e. 1-ring neighborhood) are updated in position. All boundary sample points associated with these enveloping primitives are then checked to find their new closest primitives. Since the association within the 1-ring neighborhood is updated, the post-contraction error in the 2-ring neighborhood is affected and updated. This is a local and efficient way of maintaining the affected regions of the updated primitives.
Note that the operation may be conservative at times for estimating the Hausdorff distance. However, this error overestimation is seldom; even when it happens, the robustness and the error control ability of our method are not compromised.
When the smallest post-contraction error of all edges is larger than a given user-specified threshold,
the algorithm terminates and outputs the simplified medial mesh. We summarize this simplification procedure in Algorithm~\ref{alg:simplification}.

\begin{algorithm}[htbp!]
\caption{\textbf{Medial Axis Simplification Based on Edge Contraction}}
\label{alg:simplification}
\begin{algorithmic}[1]
\STATE {\em Initialization}--Compute the post-contraction error for all edges and store them in a priority queue.
\WHILE{the smallest post-contraction error is less than a given threshold}
		\STATE Pop the edge with the smallest post-contraction error;
		\STATE Contract the edge to one of its endpoints;
		\STATE Re-evaluate the post-contraction error for edges affected by the contraction and update the priority queue.
\ENDWHILE
\STATE {\em Return} the simplified medial axis.
\end{algorithmic}
\end{algorithm}


\subsection{Homotopy Preservation}

It is an important requirement for medial mesh simplification that the topology of the boundary surface $\partial \mS$ of an input shape $\mS$ be preserved during simplification. Suppose that an edge merging step simplifies a medial mesh $\mC_1$ to the medial mesh $\mC_2$. Let $\partial \mS_1$ and $\partial \mS_2$ be the boundary surfaces of the shapes represented $\mC_1$ and $\mC_2$, respectively. Specifically, we require that $\partial \mS_1$ and $\partial \mS_2$ be isotopic, which means that (1) $\partial \mS_1$ and $\partial \mS_2$ are homomorphic; and (2) there is a continuous deformation of homomorphism from $\partial \mS_1$ to $\partial \mS_2$. To meet this requirement, we perform topological checking at both local and global levels. Locally, we make sure at each edge merging step that the topology of the local region of $S_i$ affected by a merged edge is not changed (i.e., a disk region is simplified into a disk region). This local checking on the homomorphism between affected regions before and after the merge aims to preserve surface topology around feature points, such as corners and creases.

\begin{figure}
\begin{minipage}{.33\linewidth}
\centering
\includegraphics[width=\linewidth]{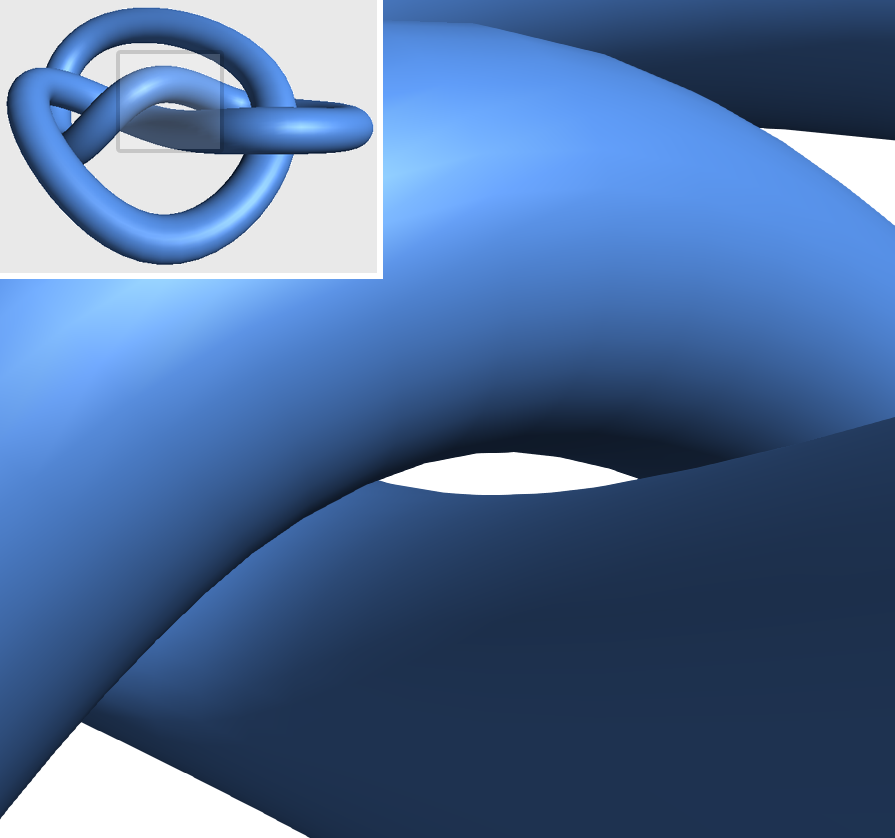}\\
\small ($a$) $\textrm{\em lfs} = 0.008$
\end{minipage}\hfill
\begin{minipage}{.33\linewidth}
\centering
\includegraphics[width=\linewidth]{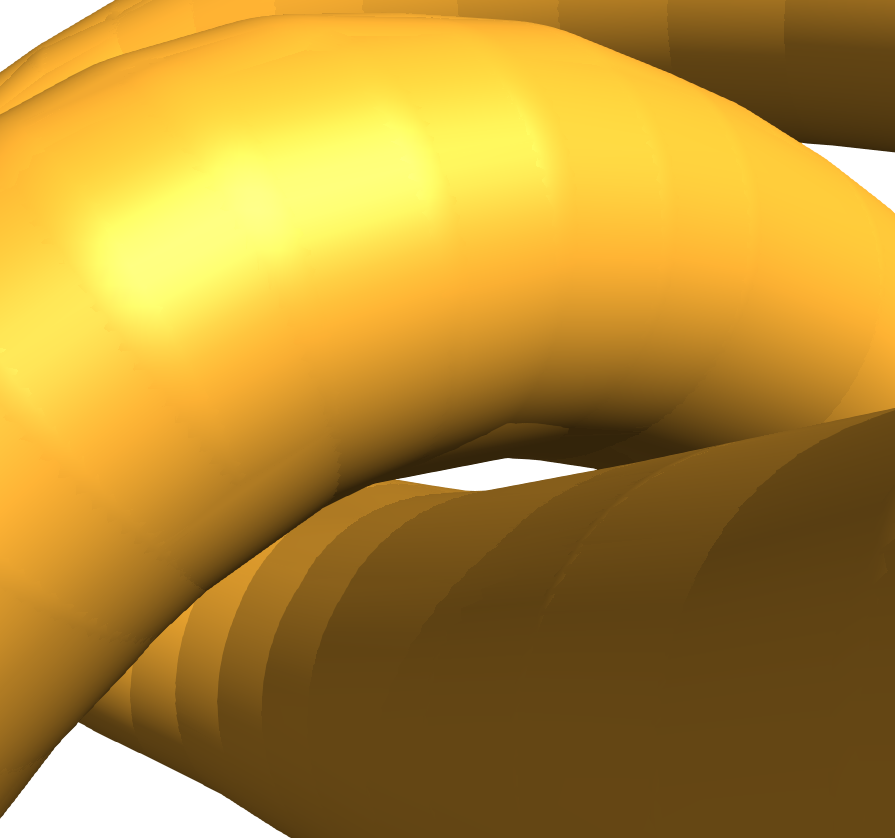}\\
\small ($b$) $\varepsilon = 0.003$
\end{minipage}\hfill
\begin{minipage}{.33\linewidth}
\centering
\includegraphics[width=\linewidth]{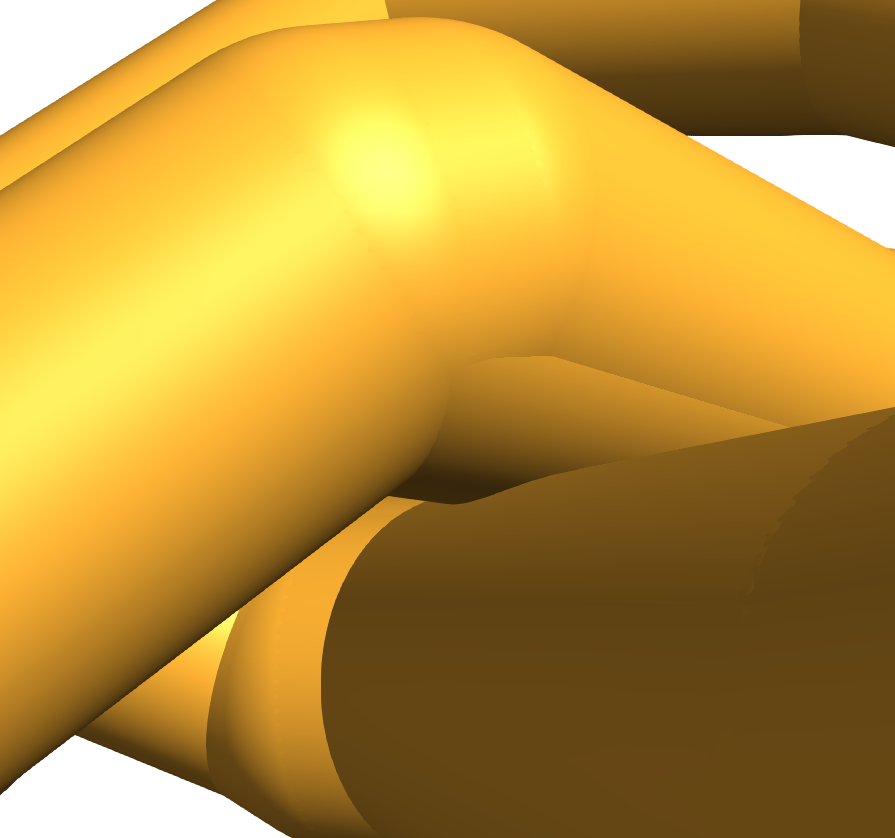}\\
\small ($c$) $\varepsilon = 0.03$
\end{minipage}
\caption{Global self-intersection prevention.  (b) Setting an error threshold that is smaller than half of the local feature size ({\em lfs}) will ensure no global self-intersection for the boundary surface represented by the simplified medial mesh. (c) Self-intersection occurs for an error tolerance larger than $\textrm{\em lfs}/2$. }\label{fig:homotopy}
\end{figure}

Note that the local topological checking is necessary but not sufficient, since the topology of the boundary surface may also change due to global self-intersection caused by edge merging, as illustrated in \autoref{fig:homotopy}. To prevent this from happening during medial mesh simplification, we first analyze the input boundary surface $\partial \mS$ to compute its local feature size ({\em lfs})~\cite{Amenta1998}. Let $\partial \mS^+_{d}$ ad $\partial \mS^-_{d}$ denote respectively the inner and outer offset surfaces of $\partial \mS$ with offset distance $d = \textrm{\em lfs}/2$. (An offset surface of $\partial \mS$ consists of points that have the constant distance $d$ to $\partial \mS$.)  Clearly, the boundary surface $\partial \mS$ lies in the volume $V_d$ bounded by the offset surfaces  $\partial \mS^+_{d}$ ad $\partial \mS^-_{d}$. We then impose $d$ as an upper bound on the approximation error tolerance used for medial mesh simplification. By enforcing this error tolerance, the boundary surface of the final simplified medial mesh is ensured to also lie inside $V_d$, hence it is free of global self-intersection.


\subsection{Special Cases}

\paragraph{Non-manifold vertices} A medial mesh is in general a non-manifold 2D mesh surface, and so special care is needed to make its topology as simple as possible, that is, not increasing the number of its non-manifold vertices and edges. Specifically, to produce a stable and simplified medial mesh, we do {\em not} allow a non-manifold vertex to merge to a neighboring manifold vertex.  However, conversely, a manifold vertex is allowed to merge to an adjacent non-manifold vertex. Furthermore, we permit non-manifold edges to merge. Of course, all these merging cases are subject to error tolerance control.

\paragraph{Ligature points} In both 2D and 3D, the medial vertices corresponding to concave regions of a shape are referred to as ligature points in the literature \cite{Macrini2008,Macrini2010}. Altering the location of ligature points easily gives rise to envelopes covering some regions outside the given shape. To resolve this issue, we also fix the ligature points and do not merge them to their neighbors. Note that we do not detect ligature points explicitly, but only fix a medial vertex when merging it with a neighboring vertex leads to a region outside the given shape with a Hausdorff distance larger than the user-specified error threshold.

\section{Experimental Results}

\begin{figure}
{\small
\centering
	\includegraphics[width=.3\linewidth]{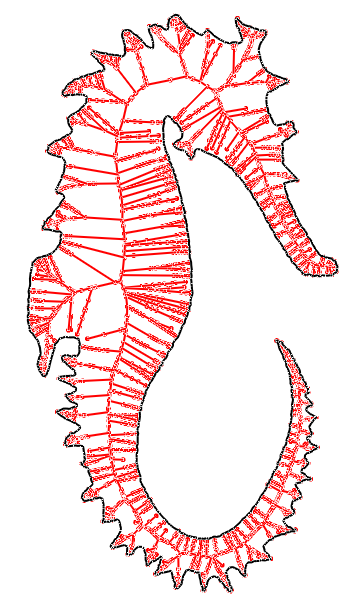}	\hfill
	\includegraphics[width=.3\linewidth]{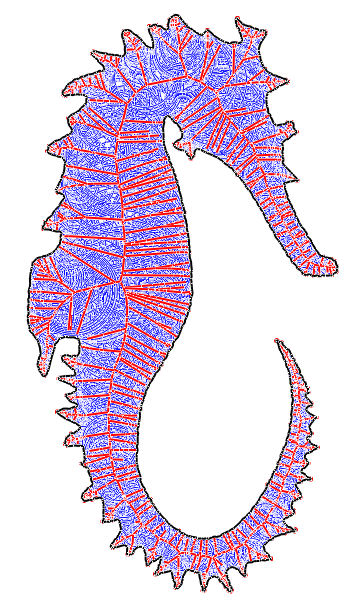}	\hfill
	\includegraphics[width=.3\linewidth]{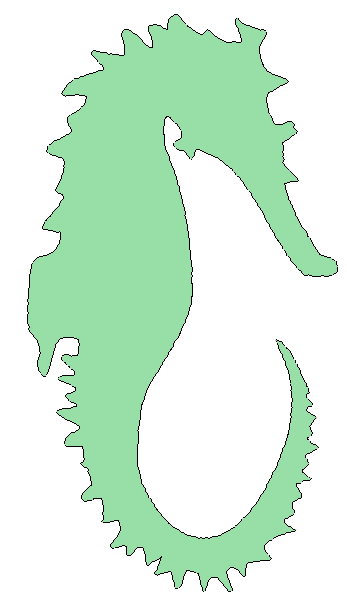}\\
	{\bf Original}: $\#$ of medial points = 5531, approximation error = $0.0$\\
	\makebox[.3\linewidth]{($a$)} \hfill
	\makebox[.3\linewidth]{($b$)} \hfill
	\makebox[.3\linewidth]{($c$)}  \\
  \vspace{1.7ex}
	\includegraphics[width=.3\linewidth]{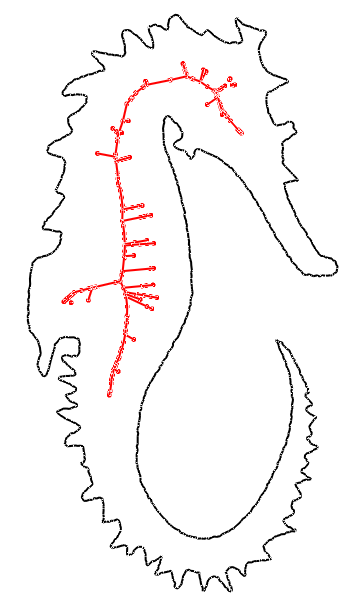} \hfill
	\includegraphics[width=.3\linewidth]{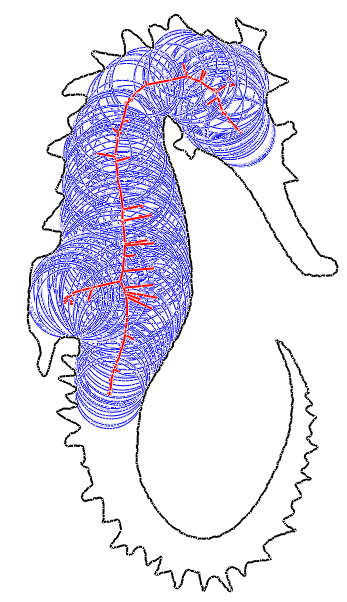} \hfill
	\includegraphics[width=.3\linewidth]{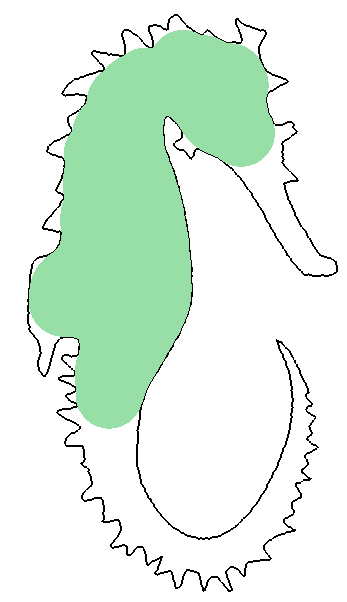}\\
	{\bf $\lambda$ medial axis}: $\#$ of medial points = 294, approximation error = $3.0e{-1}$\\
	\makebox[.3\linewidth]{($d$)} \hfill
	\makebox[.3\linewidth]{($e$)} \hfill
	\makebox[.3\linewidth]{($f$)}  \\
  \vspace{1.7ex}
	\includegraphics[width=.3\linewidth]{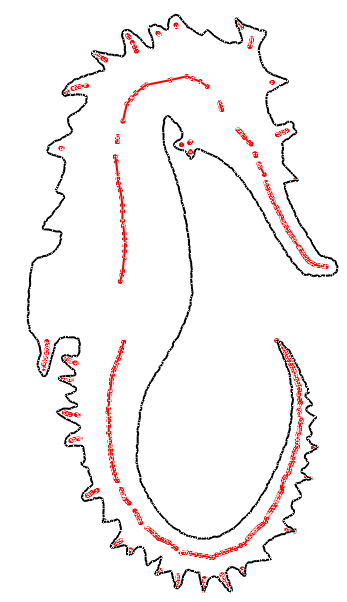} \hfill
	\includegraphics[width=.3\linewidth]{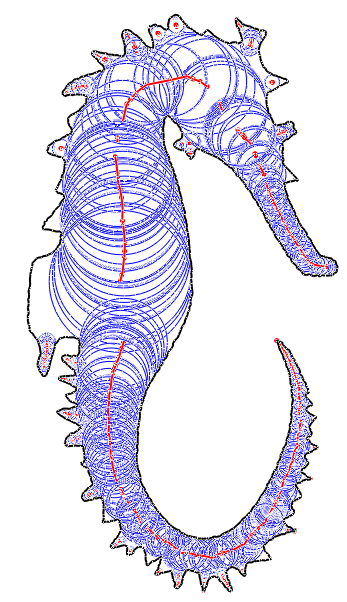} \hfill
	\includegraphics[width=.3\linewidth]{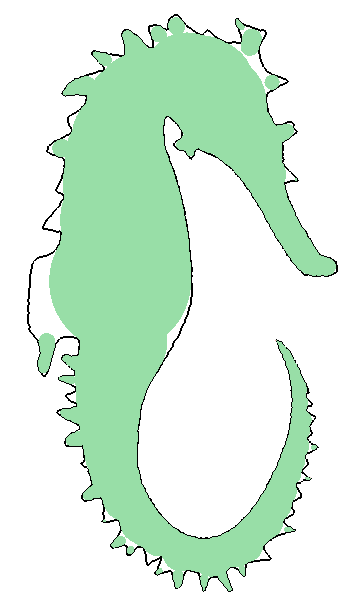}\\
	{\bf Angle-based}: $\#$ of medial points = 1429, approximation error = $3.9e{-2}$\\
	\makebox[.3\linewidth]{($g$)} \hfill
	\makebox[.3\linewidth]{($h$)} \hfill
	\makebox[.3\linewidth]{($i$)}  \\
  \vspace{1.7ex}
	\includegraphics[width=.3\linewidth]{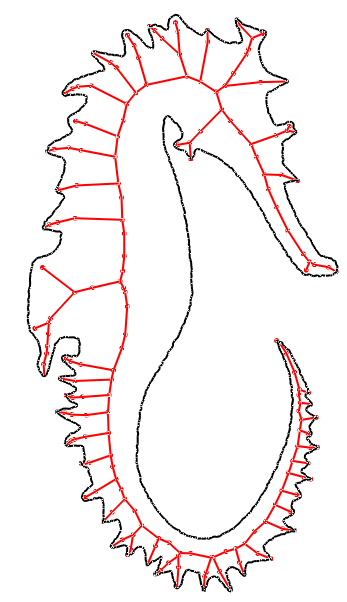} \hfill
	\includegraphics[width=.3\linewidth]{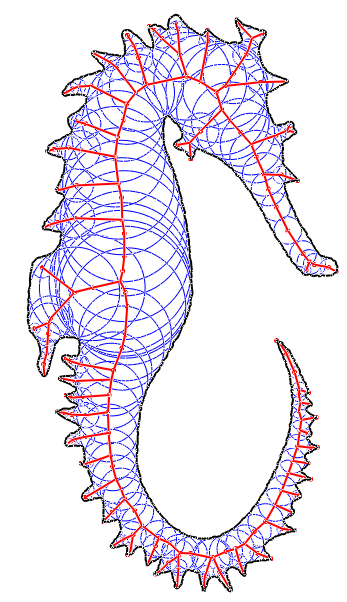} \hfill
	\includegraphics[width=.3\linewidth]{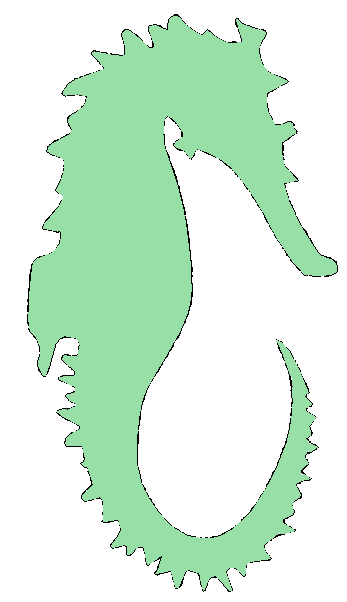}\\
	{\bf Medial mesh}: $\#$ of medial points = 194, approximation error = $1.6e{-3}$\\
	\makebox[.3\linewidth]{($j$)} \hfill
	\makebox[.3\linewidth]{($k$)} \hfill
	\makebox[.3\linewidth]{($l$)}  \\
} 
	\caption{A comparison of medial axis simplification methods. The first column shows the original medial axis, the filtered medial axis by the $\lambda$ medial axis method, the angle-based method and our method, respectively. The second column shows these medial axes together with medial circles. The third column shows the reconstructed shape boundary by these different methods.}
    \label{fig:2d_unstable}
\end{figure}

In this section we present the testing of our method and its comparison with several existing methods on 3D models of a wide variety of shapes, complexities and topologies. All experiments are performed on a Windows $7$ workstation with an Intel i7 CPU and 12 GB main memory. While a method such the one~\cite{Amenta1998} provides a filtered medial axis which we could use as input, to  demonstrate the simplification ability of our method, we choose to use the Voronoi diagram of sampled points on the input boundary surface as the initial medial axis which is typically highly unstable. We compute the Voronoi diagram using the CGAL package {\em Delaunay Triangulation 3}~\cite{cgal:pt-t3-12b} to compute the Delaunay triangulation and then taking its dual. The conversion from a medial mesh to a triangle mesh is carried out using the CGAL package {\em Skin Mesh Generation}~\cite{cgal:k-ssm3-12b}.
Our algorithm is fast and has small memory footprint. For a typical 3D model with 4K vertices, our algorithm generates a simplified medial mesh with approximation error 0.001 (relative the normalized diagonal of the bounding boxes of the input shape) using around 20 seconds with 130MB memory usage.


%
%

\paragraph{Comparison with criterion based methods}
We first compare our method to two existing medial axis pruning methods, the $\lambda$ medial axis~\cite{Chazal2005} and the angle-based~\cite{Foskey2003} methods, using a 2D seahorse shape shown in \autoref{fig:2d_unstable}.
The $\lambda$ medial axis method performs filtering using the circumradius criterion and completely removes the head and tail of the seahorse while the noise on the trunk still remains, largely due to the different feature scales of the input shape (\autoref{fig:2d_unstable}(d)-(f)).
The angle-based method, on the other hand, does not preserve the input topology and the resulting medial axis simply becomes  disconnected (\autoref{fig:2d_unstable}(g)-(i)).
In addition, a main branch on the left has been incorrectly removed, therefore compromising significantly the approximation accuracy of the pruned medial axis.
The medial mesh computed by our method, in contrast, uses much fewer sample circles while achieving the smallest approximation error among all. Visually, the union of envelopes of adjacent medial circles in the medial mesh yields a more accurate shape approximation, as shown in \autoref{fig:2d_unstable}(l). In comparison, gaps between the union of medial circles and the original shape boundary are clearly visible (\autoref{fig:2d_unstable}(i)).
We also tested a variant of the angle-based method by enforcing topology preservation during filtering. It uses 2,978 medial points to achieve an approximation error of 0.01, while a media mesh uses only 79 primitives for the same error level.

%

\begin{figure*}[t!]
\includegraphics[width=\linewidth]{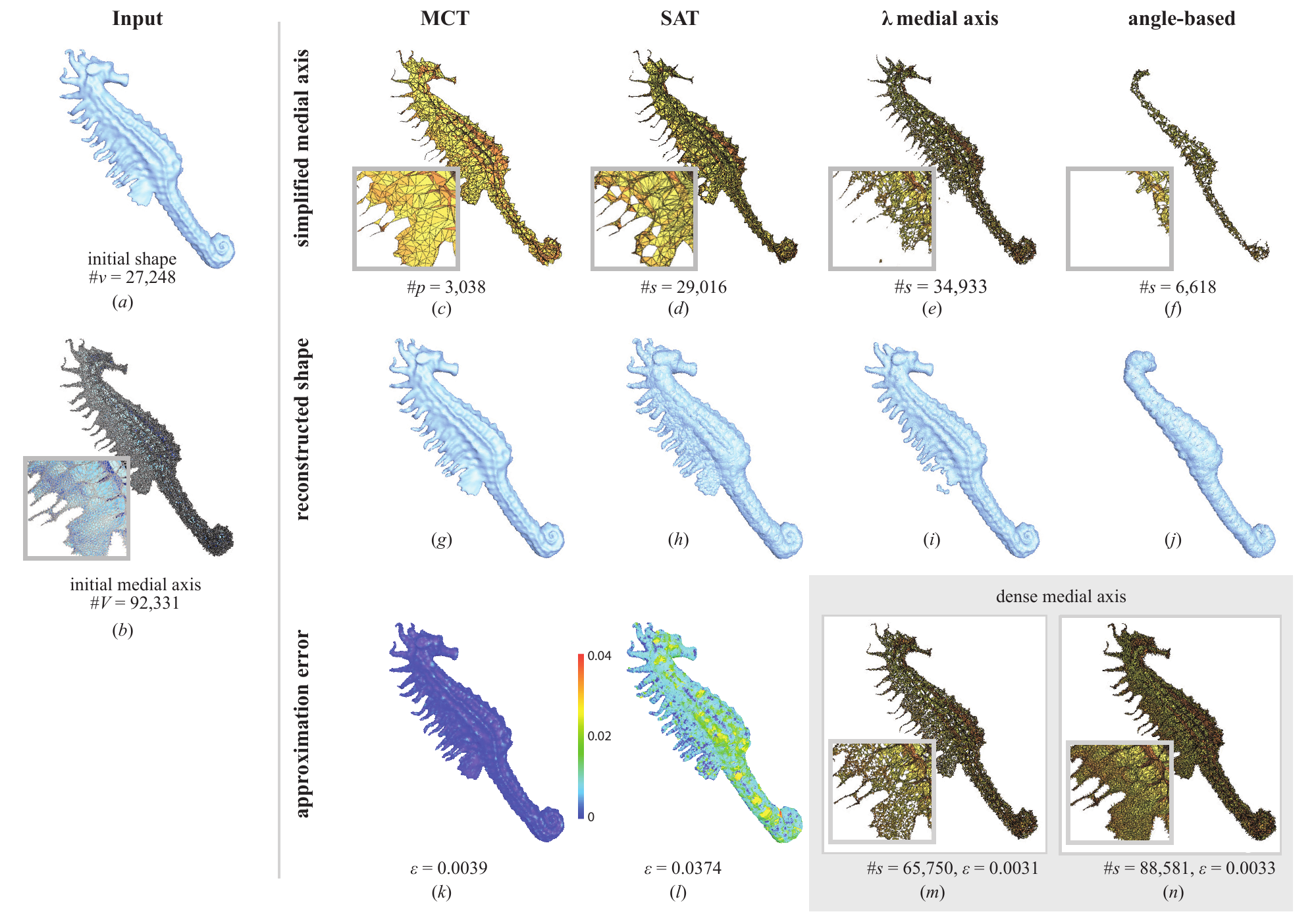}
\caption{Comparisons of 3D medial axis simplification methods. The first column ($a$ and $b$) shows the original seahorse model and the initial medial axis. The first row ($c$-$g$) shows the medial axes computed by different methods. The second row ($h$-$l$) shows the reconstructed boundary surfaces from the respective medial axes in the first row.
($m$) and ($n$) show the color-coded approximations of the reconstructed surfaces in ($h$) and ($i$), respectively. ($o$) and ($p$) are the dense and instable medial axes generated by the $\lambda$ medial axis method and the angle-based method, respectively.}
\label{fig:seahorse_3d}
\end{figure*}

We next use a 3D shape to compare our method with the SAT method~\cite{Miklos2010}, in addition to the $\lambda$ medial axis method and the angle-based method, as shown in~\autoref{fig:seahorse_3d}. This 3D seahorse model has 27K vertices and its initial medial axis has many unstable spikes (\autoref{fig:seahorse_3d}(b)). Our method significantly simplifies the medial axis to reduce the number of primitives to $\sim$3K, with this highly simplified medial mesh still achieving an accurate shape reconstruction with approximation error smaller than $0.004$ only (\autoref{fig:seahorse_3d}(c,g,k)).
In contrast, the shape approximation error of the filtered medial axis from SAT is 0.03741 even with
$\sim$30K spheres, about an order of magnitude more primitives than our method.
Moreover, the topology of the medial axis has not be preserved (e.g., the holes created at the lower
backbone of the seahorse in~\autoref{fig:seahorse_3d}(d,h,l)).
Clearly, medial axis filtering with both $\lambda$ medial axis and angle-based methods
led to an unacceptable shape reconstruction or topological change (\autoref{fig:seahorse_3d}(e,i,f,j)). Meanwhile, the number of spheres used by either method is at least an order of magnitude larger than that in our method.
To attain the same shape approximation errors as our result, a huge number of spheres would be necessary with both methods, with the corresponding medial axes being highly unstable (\autoref{fig:seahorse_3d}(m,n)).

Admittedly, these three methods (SAT, $\lambda$ medial axis and angle-based methods) are mainly designed for removing medial axis instability, rather than specifically for shape approximation.
Nevertheless, our simplification method based on the medial mesh achieves both accurate shape approximation and a stable medial axis representation simultaneously.
This is due to the fact that medial axis instabilities correspond to small perturbations on the shape boundary, the removal of which would not incur a significant shape approximation error and, therefore, they can be effectively eliminated with our error-driven simplification.

We also examine the simplification power of our method irrespective to the use of an enveloping representation.  To this end, we compute simplified medial axes using the other three methods, and build the same enveloping representation from their resulting medial vertices just like the medial mesh.  \autoref{tab:all_methods_envelope} shows that the medial mesh still uses far fewest number of primitives to attain the same approximation error among all four medial axis simplification methods.

\begin{table}
\centering
\begin{tabular}{c|ccc}
          & \# medial vertices & \# primitives & error \\ \hline
 Medial mesh     &  352    & 2,586        &  0.02219 \\
 SAT    &  8,131    & 50,161        &  0.02316 \\
$\lambda$ medial axis & 58,221 & 355,672 & 0.02293 \\
angle-based & 48,158 & 300,632 & 0.02453
\end{tabular}
\caption{Number of primitives used for representing the 3D seahorse model, with all four methods using the enveloping representation like a medial mesh.
}\label{tab:all_methods_envelope}
\end{table}

Although SAT may often result in topologically incorrect shape reconstruction, an issue acknowledged in~\cite{Miklos2010}, it generally produces good shape approximation, though with a large number of vertices. In comparison, our method achieves the same approximation precision using far fewer number of primitives than SAT, while preserving shape topology. This is important for many applications such as shadow computation and shape deformation where the computational complexity depends on the number of primitives. \autoref{tab:MCT vs SAT} shows a comparison of our method against SAT on several 3D models in terms of the number of primitives used against the approximation precision that can be achieved.
It can be seen that for high-precision approximation ($\varepsilon=0.001$), SAT requires at least two orders of magnitude more primitives than a simplified medial mesh.
Even for low-precision approximation ($\varepsilon=0.032$), the number of primitives needed by SAT is still more than one order of magnitude larger than the medial mesh.

\begin{table}[th!]
\resizebox{\linewidth}{!}{
\begin{tabular}{c| c | c c c c c c}
\multicolumn{2}{c|}{} & \multicolumn{6}{c}{approximation error}\\ \cline{3-8}
\multicolumn{2}{c|}{} & $0.032$ & $0.016$ & $0.008$ & $0.004$ & $0.002$ & $0.001$\\
	\hline\hline
{\em retinal}&   SAT  & $8075$ & $27251$ & $89378$ & $277990$ & $827632$ & $1928799$\\
   & Medial mesh  & $178$ & $294$ & $498$ & $855$ & $1625$ & $3035$\\\hline
{\em bird} & SAT  & $6785$ & $25365$ & $98771$ & $383880$ & $1392083$ & $5086712$\\
    & Medial mesh  & $265$ & $478$ & $839$ & $1479$ & $2296$ & $5503$\\ \hline
{\em	table} & SAT  & $12833$ & $52463$ & $202814$ & $773646$ & $2971080$ & $11542309$\\
  &  Medial mesh  & $116$ & $327$ & $886$ & $2378$ & $5645$ & $10759$\\ \hline
{\em girl} &  SAT  & $11625$ & $37701$ & $127557$ & $430056$ & $1276757$ & $3245069$\\
  & Medial mesh  & $506$ & $776$ & $1476$ & $2652$ & $3913$ & $7128$\\ \hline
{\em fandisk} & SAT  & $19344$ & $71818$ & $273847$ & $1047419$ & $3875252$ & $13130365$\\
  & Medial mesh & $242$ & $398$ & $587$ & $904$ & $1742$ & $3059$\\ \hline
\end{tabular}
} 
\caption{Comparisons of the approximation error against the number of primitives in the simplified medial axis for the SAT and our methods on five 3D models.}
\label{tab:MCT vs SAT}
\end{table}

\begin{figure}
\includegraphics[width=\linewidth]{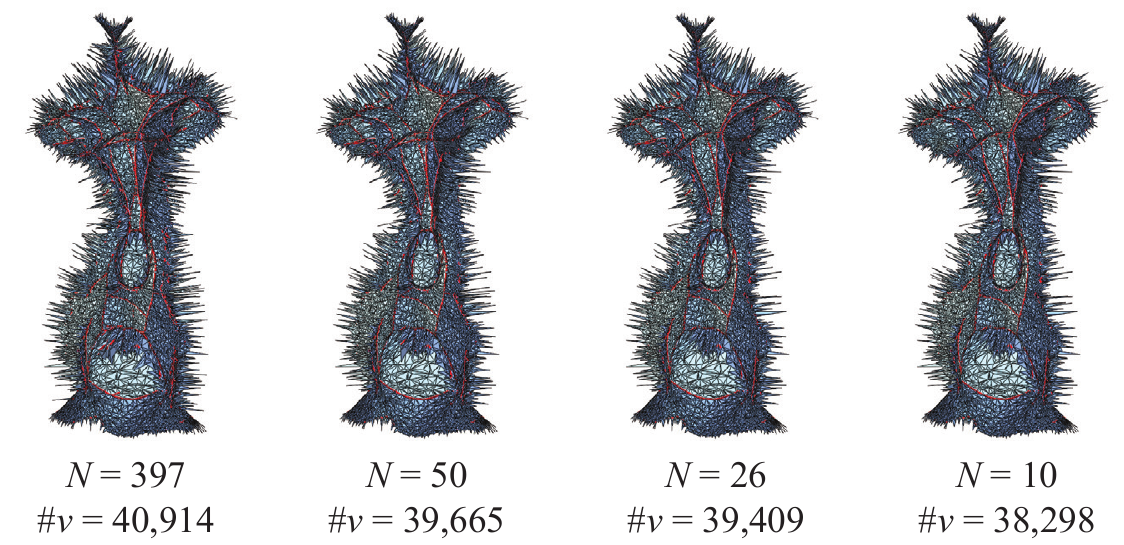}
\caption{The feature-based method Tam and Heidrich [2002] progressively reduces the number of medial sheets ($N$) to achieve shape simplification, taking a noisy Voronoi diagram of the shape as the initial input.  The instability of the medial axis and the number of vertices are not reduced significantly as the medial sheets are gradually removed.  The non-manifold edges are marked in red.}\label{fig:tam}
\end{figure}

\paragraph{Comparison with feature-based methods}
We now compare with another medial axis simplification method based feature sheets by Tam and Heidrich~\shortcite{Tam2002}. An initial axis surface is first segmented by this method into a collection of sheets which are maximal manifold patches intended to represent the features or components of the original shape. Then the medial axis is simplified by deleting these sheets one by one in an increasing order of volume errors incurred by deleting the sheets. Using a small volume threshold, the method is able to remove insignificant features while keeping relatively large sheets of medial axis representing important parts of the original shape. If the volume error threshold increases, significant pieces of the shape will begin to be trimmed off.  In any case, the remaining sheets are still a dense representation containing a large number of triangles, as shown in our comparison results in \autoref{fig:tam}. Therefore, this method is not effective at all for the purpose of geometric simplification; rather, it is primarily for simplifying the structure of a medial axis by removing those parts that are insignificant for representing the boundary. Our method, on the other hand, computes an accurate yet compact simplified representation.

\begin{figure}
\includegraphics[width=\linewidth]{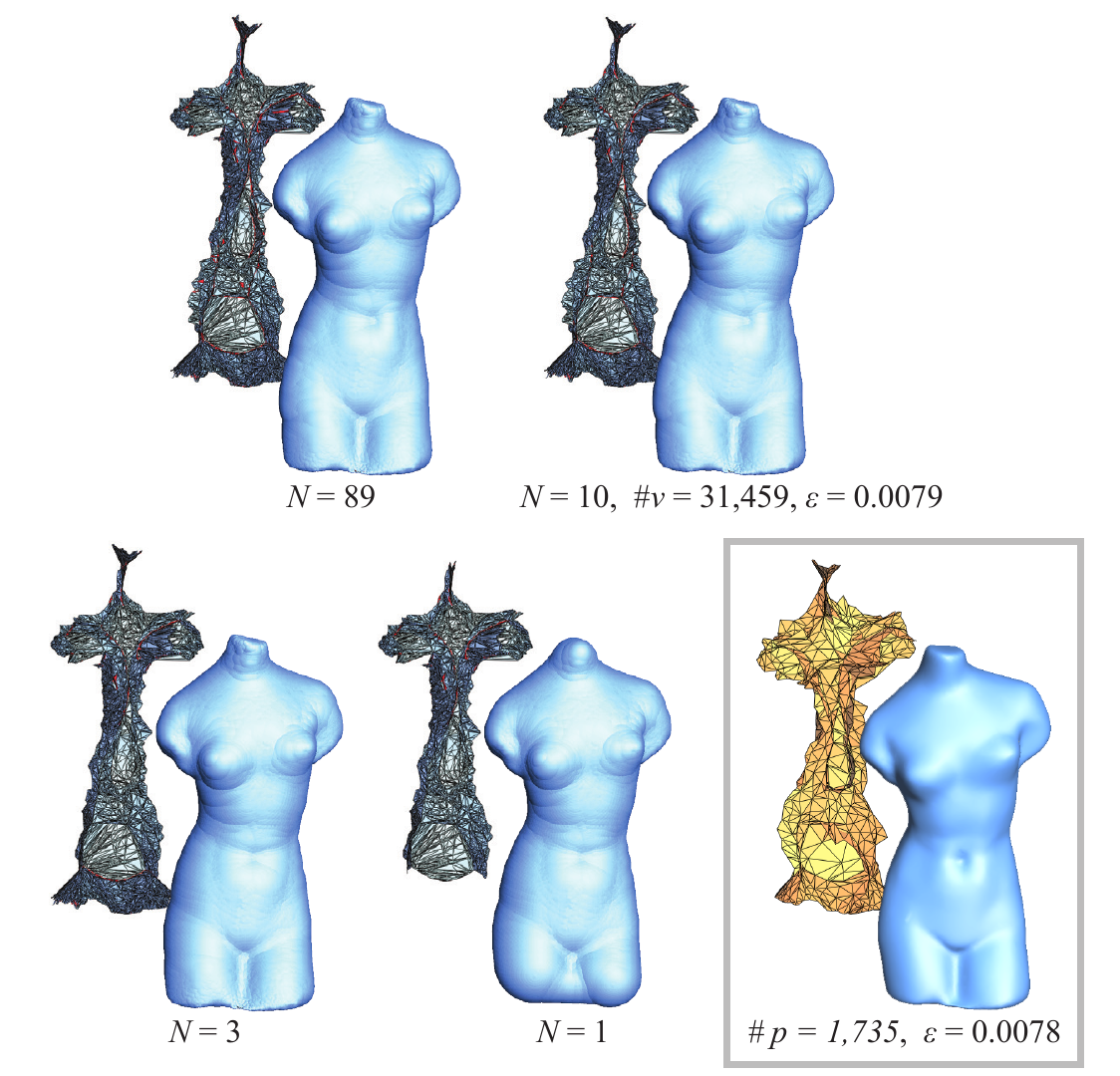}
\caption{The feature-based method Tam and Heidrich [2002] progressively reduces the number of medial sheets ($N$) to achieve shape simplification, taking a filtered medial axis as the initial input.  Our method achieves the same approximation accuracy at $\varepsilon=0.0078$ with much fewer number of primitives used.}\label{fig:tam_sat}
\end{figure}

It is worth noting that the method of Tam and Heidrich needs a relatively clean medial axis as its initial medial axis, which can be provided by Power Crust, for example, as mentioned in~\cite{Tam2002}. Otherwise, if initialized with a highly unstable unfiltered medial axis from the Voronoi diagram, this method is often unable to prune unstable spikes that are not separate sheets, as shown in \autoref{fig:tam_sat}. Hence, in this sense, this method is not a complete simplification method by itself.

\begin{figure}
\includegraphics[width=\linewidth]{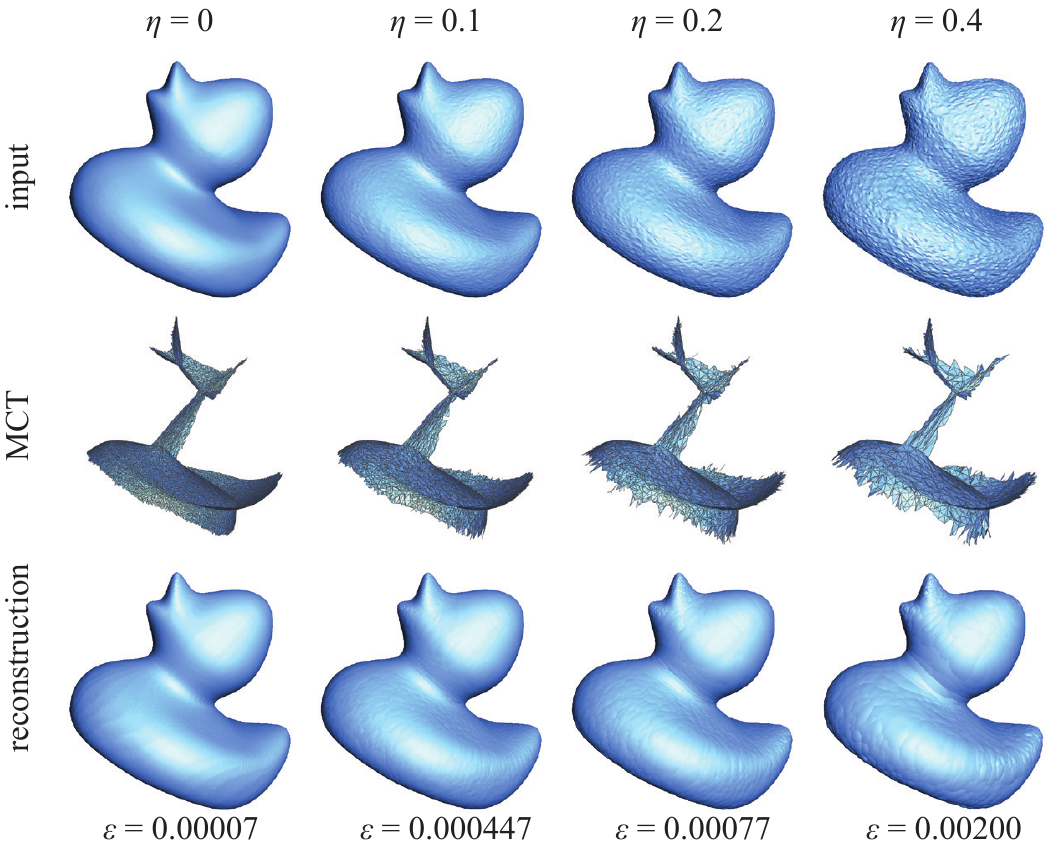}
\caption{Medial meshes for a noisy model on different level of noise.  The topology of the medial meshes is resistant to noisy input.  Noises are removed effectively from the accurate reconstruction.  ($\varepsilon$ is the Hausdorff distance between the reconstructed surface and the noise-free input.) }\label{fig:noise-duck}
\end{figure}

\paragraph{Noisy models}
As MAT is in general sensitive to shapes with noisy boundaries,
we would like to investigate how the medial mesh simplification is affected by noisy models.
We apply different level of noise relative to the average edge length of an input model (measured by $\eta \in [0,1]$) and compute the corresponding simplified medial mesh.  The result is shown in \autoref{fig:noise-duck}.
The medial mesh simplification is capable of obtaining a stable medial representation and at the same time removing noise effectively as much as to a noise level of $\eta = 0.2$.  The medial mesh retains very similar topology despite the increasing noise level.
Also, it faithfully reproduces the original unnoisy input as indicated by the small distance error of the reconstruction.

%
%
%
%

\paragraph{Volume simplification}

\begin{figure*}[!ht]
\includegraphics[width=\linewidth]{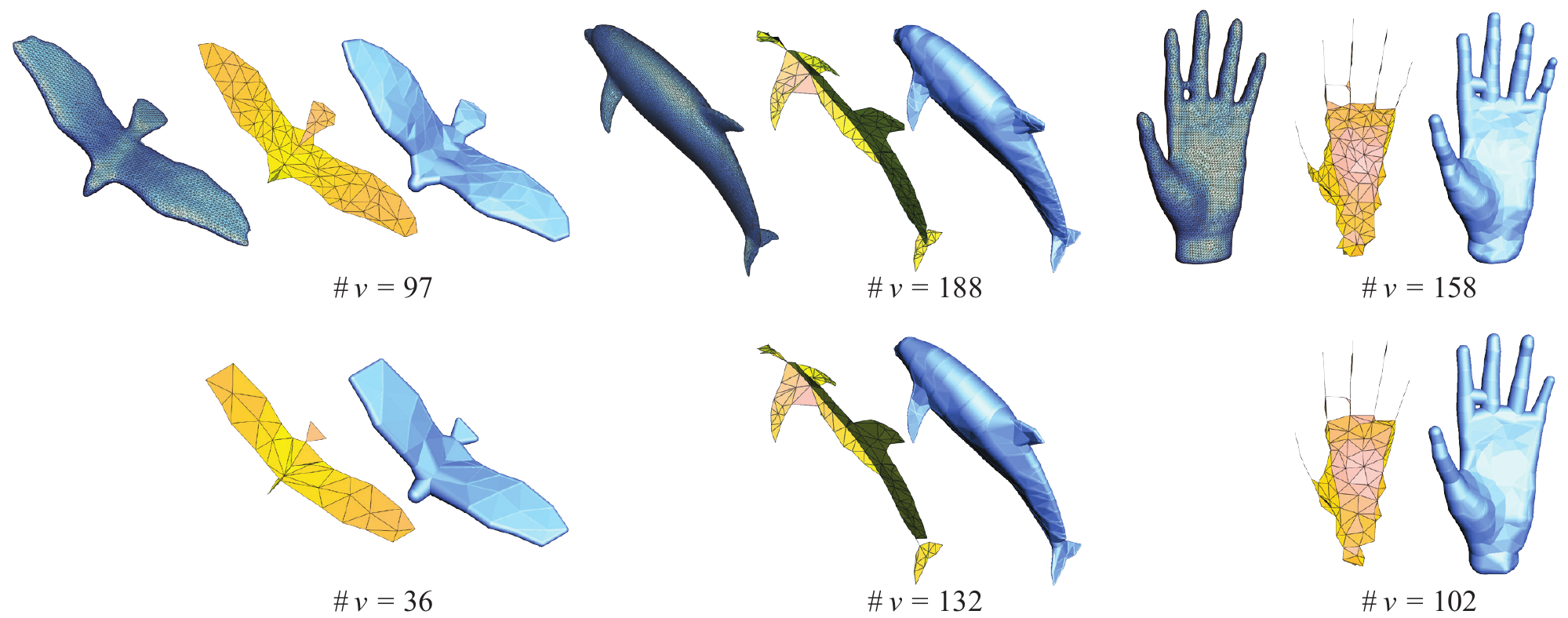}
\caption{Extreme simplification of shapes.  Our method is capable of generating medial meshes with very few medial vertices ($\#v$), yet preserving the volume features of the original shapes well.}\label{fig:extreme}
\end{figure*}

Figure~\ref{fig:teaser} shows that our method is capable of generating a series of volume approximation at progressive levels of simplification. Note that the basic shape of an object is still preserved even when only a small number of medial spheres are kept. Figure~\ref{fig:extreme} shows three other examples computed by our method.  It can be seen that even at such extreme simplification levels, volume features of the original shape can still be retained. This property is highly desirable in applications such as shape analysis and progressive shape transmission. 

\section{Applications}

\subsection{Shape Approximation}

\begin{figure}[ht!]
\begin{minipage}{0.49\linewidth}
\centering
\includegraphics[width=\linewidth]{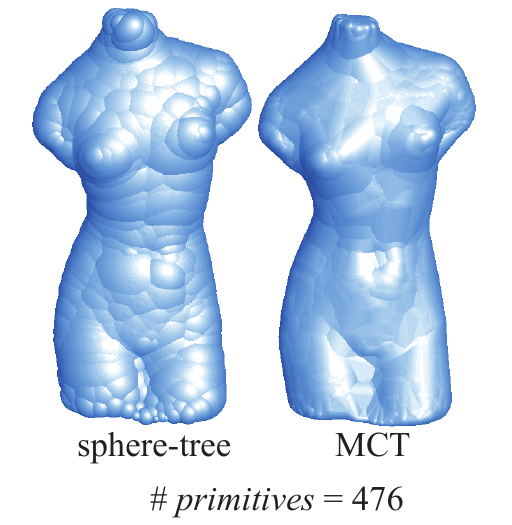}\\
({\em a})
\end{minipage}\hfill
\begin{minipage}{0.49\linewidth}
\centering
\includegraphics[width=\linewidth]{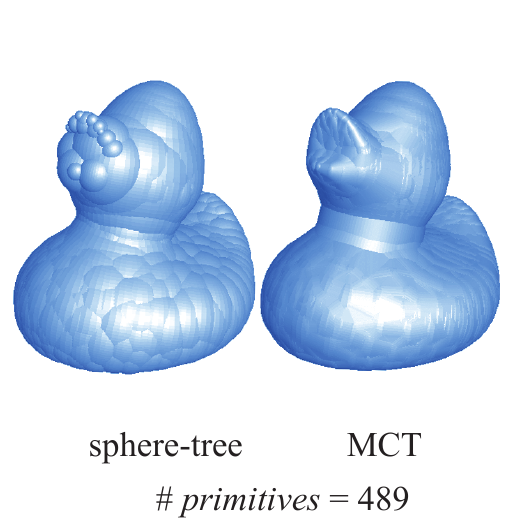}\\
({\em b})
\end{minipage}
\caption{Two models, (a) Venus, and (b) Duck, used for comparison of shape approximation
in \autoref{tab: shape_approx}.
The same number of primitives are used for both methods in each case.
}\label{fig:shape_approx}
\end{figure}

\begin{table}
\centering
\begin{tabular}{c|ccc}
          & sphere-tree & medial spheres & Medial mesh \\ \hline
 Duck     &  $>$ 0.050    & = 0.029        &  = 0.004 \\
 Venus    &  $>$ 0.054    & = 0.033        &  = 0.004 \\
 Bunny    &  $>$ 0.091    & = 0.062        &  = 0.007
\end{tabular}
\caption{Volume difference with respect to the original shape.  The same 500 number of primitives are used for all 3 methods and all 3 models. Data for the sphere-tree and medial spheres methods are taken from Table 1 of \protect\cite{Stolpner2011a}.
}\label{tab: shape_approx}
\end{table}

The medial mesh proposed in this paper provides an effective alternative to shape approximation, compared to the conventional spheres representations~\cite{Bradshaw2004,Stolpner2011a}. The key difference from the previous methods is that we use the interpolation of the medial spheres to approximate a given shape; in other words, the given shape is approximated with the union of the enveloping primitives defined by the medial mesh. Specifically, given any 3D shape to be approximated and an error tolerance provided by the user, we run our method for computing and simplifying the medial mesh. Once the simplified medial mesh is obtained, we collect all the enveloping primitives defined by the triangle faces of the mesh as well as those primitives defined by medial edges if the edges are not contained in any face. Then the union of these primitives are used as an approximation of the given shape, meeting the specified error tolerance.

Due to the use of simple sphere interpolation, the new approximation enabled by the simplified medial mesh we compute is much more efficient than the previous methods based on the union of spheres, as shown by the comparisons in \autoref{tab: shape_approx} and \autoref{fig:shape_approx}. Typically, using the same number of primitives, the approximation error of the medial mesh is one order of magnitude smaller than those of the sphere-tree~\cite{Bradshaw2004} and the medial spheres~\cite{Stolpner2011a} methods, as shown in \autoref{tab: shape_approx}.


The enveloping primitives used in medial meshes are a new type of simple primitives capable of efficient and compact shape approximation, and have potential to benefit important applications in graphics and geometry processing, such as fast shadow computation~\cite{Ren2006,Wang2006} and collision detection or avoidance~\cite{Larsen1999}.
While spheres are by themselves very simple primitives, further research would be needed of fast geometric computation involving the enveloping primitives used by the medial mesh.

\subsection{Shape Deformation}
Medial-based deformation has been a potential application of the medial axis since the very beginning~\cite{Blum1973}. Since medial axis is equipped with a radius function, its use in a deformation technique leads naturally to thickness-preserving shape deformations.  However, existing medial-based deformation techniques often take a simplified form, such as a stick skeleton, which can be viewed as a crude approximation to the medial axis of a shape.
This is largely due to the fact that there has not been handy method for obtaining a general stable medial axis with not only stick skeleton but also sheets.  The simplicity of a medial axis also benefits greatly the efficiency of a deformation scheme whose complexity generally depend on that of the medial axis or skeleton used. As such, we reckon that the simplified medial mesh generated by our method provides a practical medial structure for use in deformation applications.
%
Here, we demonstrate how to couple the medial mesh with the embedded deformation technique by Sumner et al.~\shortcite{Sumner2007} to achieve this goal. The technique in \cite{Sumner2007} lets users directly manipulate `handles' on an object and a subspace defined by a relatively sparse graph embedded inside the object is deformed by nonlinear minimization, whose energy function measures how much local transformations at the graph nodes deviate from rigid transformations. Similar to SSD, any point elsewhere on the object is under the influence of a subset of graph nodes, and its deformed position is a weighted average of the positions predicted by the transformations associated with these graph nodes.

\paragraph{Medial-based embedded deformation}
Let us now describe the use of a medial mesh as an embedded graph for deformation.
Each vertex in the medial mesh serves as a node of an embedded graph $\cG$ and is associated with a rigid transform.
The nodes of influence, $\cN_i$, for a mesh vertex $\mv_i$ includes the set of nodes $\cV_i$ forming $\mv_i$'s associated enveloping primitive as well as the 1-ring neighbors of those nodes in $\cV_i$.
The weights for vertex-node pairs are assigned by taking into account the distance from $\mv_i$ to the nodes in $\cV_i$, and the discrete geodesic distance on the medial axis to the other nodes not in $\cV_i$.


Each edge of the medial mesh remains an edge in $\cG$.
Since the medial vertices are only sparsely connected, in order to better maintain transformation consistency among nearby nodes, we augment the edge set of $\cG$ by including additional edges that connect 2-ring neighboring vertices in the medial mesh, as well as edges that connect nodes associated with a common mesh vertex.
There may also be dangling edges in the medial mesh representing some tubular regions on the input shape. The local transform along a dangling edge has ambiguity because the rotation around the edge is unspecified. Therefore, we introduce an extra energy term for regularizing twisting along dangling edges,
requiring that the difference between the rotations at the two ends of a dangling edge be minimal.

\paragraph{Thickness preservation}
Deformed vertex positions follows vertex blending in SSD and can be computed very efficiently.
Vertex blending in general does not respect the radius function of a medial mesh, which means that a mesh vertex originally on its envelope may no longer lie on the envelope of the deformed medial mesh.
A simple treatment for thickness preservation of the deformed shape is to project the deformed vertices onto the envelope of the deformed medial mesh, the latter of which can be done efficiently by considering only the shortest distances to the enveloping primitives that a vertex is initially associated with and its 1-ring neighboring enveloping primitives.
An offset distance to the envelope of the medial mesh is kept for each vertex and is restored on the deformed envelope after deformation.


\paragraph{Results}

\begin{figure}[t!]
\begin{minipage}{.48\linewidth}
\centering
\includegraphics[width=\linewidth]{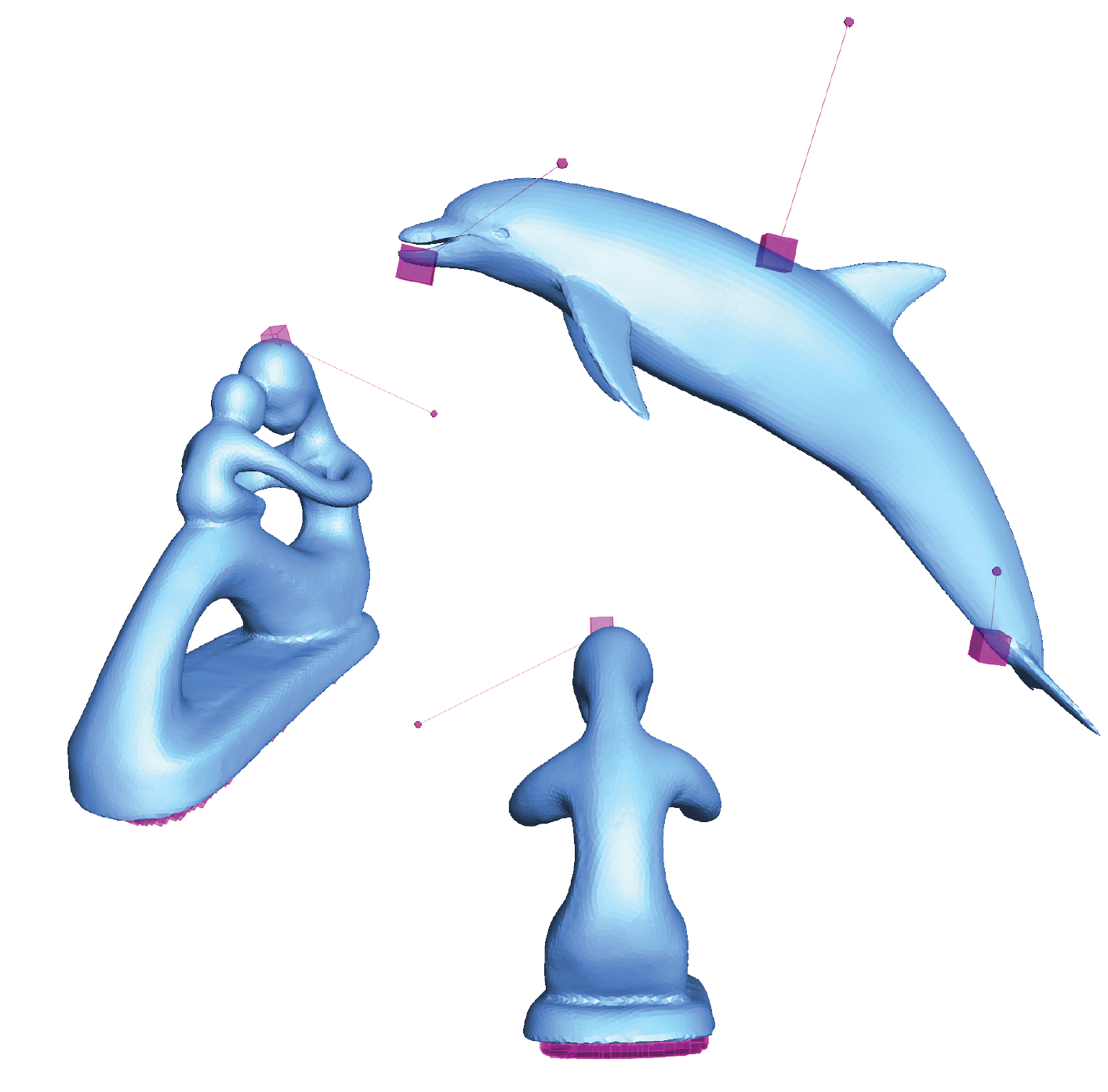}\\
\small original shape\\
\small ({\em a})
\end{minipage}\hfill
\begin{minipage}{.48\linewidth}
\centering
\includegraphics[width=\linewidth]{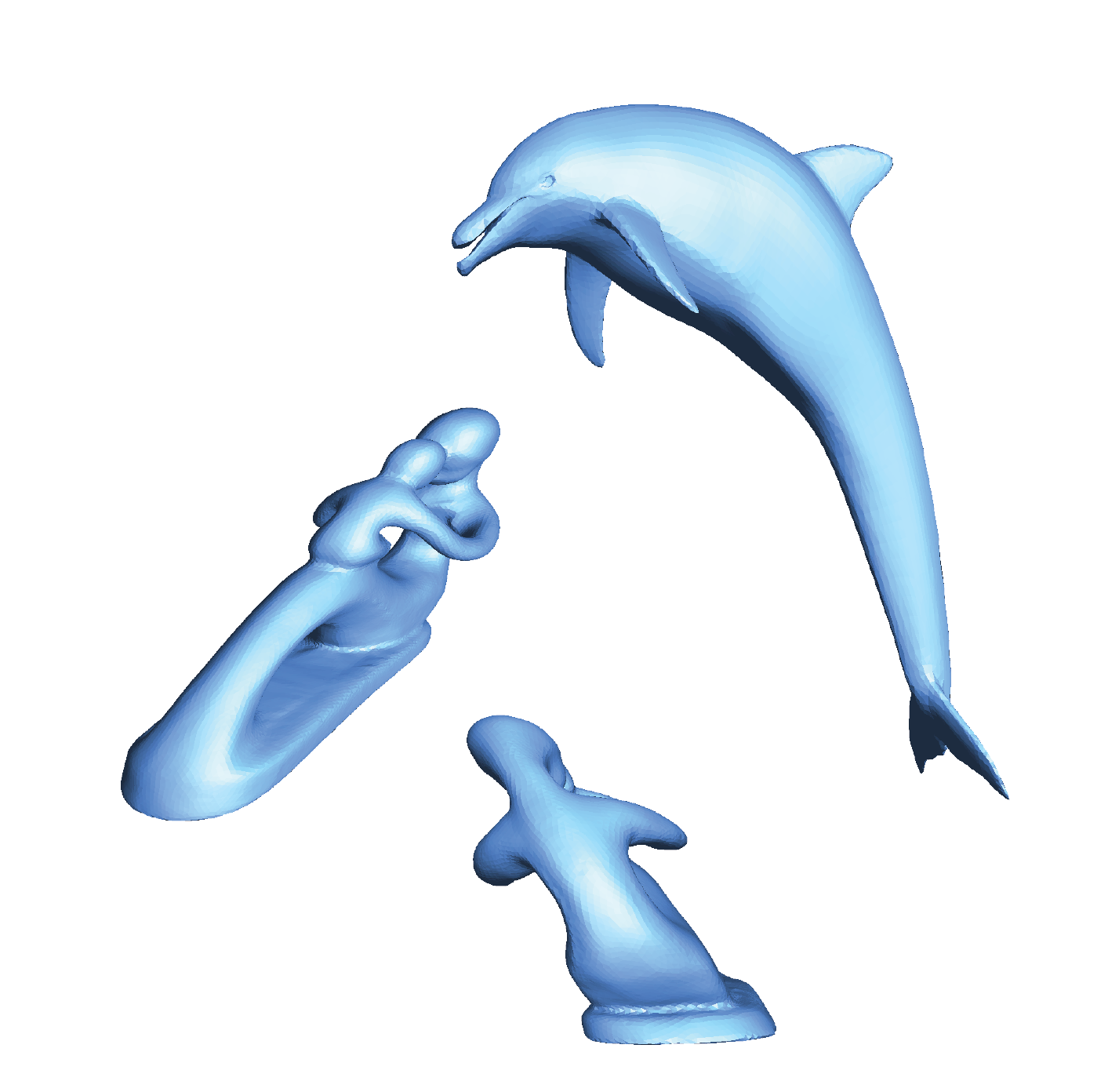}\\
\small surface graph\\
\small ({\em b})
\end{minipage}\\
\begin{minipage}{.48\linewidth}
\centering
\includegraphics[width=\linewidth]{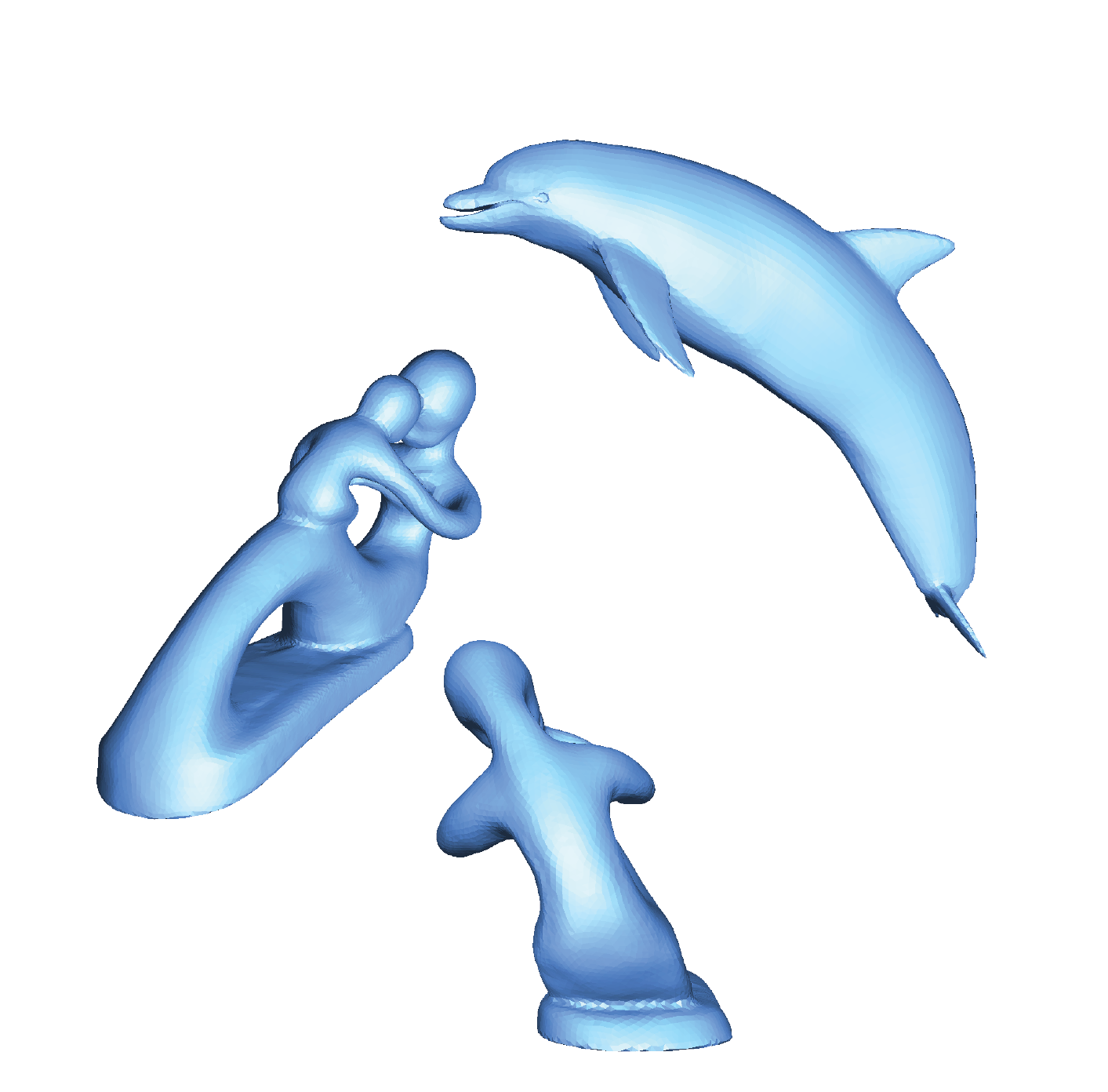}\\
\small volume graph\\
\small ({\em c})
\end{minipage}\hfill
\begin{minipage}{.48\linewidth}
\centering
\includegraphics[width=\linewidth]{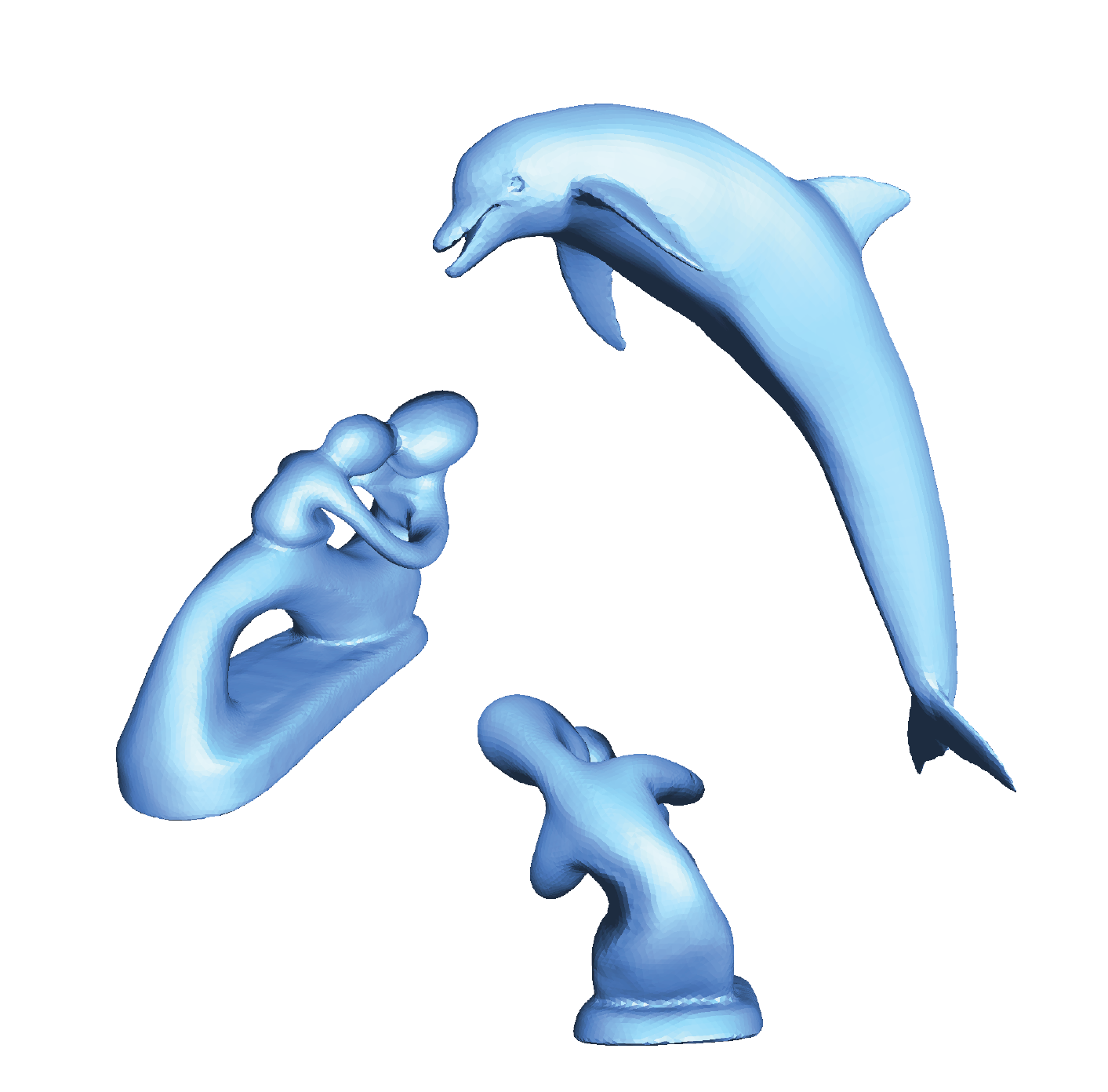}\\
\small Medial mesh\\
\small ({\em d})
\end{minipage}
\caption{Embedded deformation of the dolphin and fertility models using different embedded graphs.
The red boxes in (a) are the point handles for defining the user specified constraints and the corresponding red dots are the intended deformed positions.}
\label{fig: deformation}
\end{figure}

We have applied the medial-based embedded deformation to various mesh models and
two results are shown in \autoref{fig: deformation}.
In addition, we have also compared our results with two other variants based on embedded graphs.
In the first variant, we use a surface graph originally suggested in~\cite{Sumner2007} for an input mesh. In this surface graph, nodes are evenly distributed samples drawn on the mesh surface. Each mesh vertex is associated with its $k$-nearest nodes ($k=4$) in the graph, and edges of the graph connect every two nodes that have a common association with a mesh vertex.
In the second variant, we enhance the surface graph with a volume graph by adapting the volumetric graph Laplacian technique by Zhou et al.~\shortcite{Zhou2005}. We first generate a dense tetrahedral mesh $\cV_m$ that retains the vertices and edges in the input surface mesh. A tetrahedral mesh $\cV$ at a much coarser level is constructed and used as an embedded volume graph.
The local transform at the vertices of $\cV_m$ is derived from the rotations of its $4$-nearest nodes in $\cV$ and is applied to the per-vertex Laplacian; the resulting rotated Laplacians are then used for reconstructing the deformed mesh.

%
%
%
%

\autoref{fig: deformation}(b) shows the result using the surface graph.
For the dolphin model, since the graph nodes at the back and on the tummy of the mesh object are not closely connected, the surface at these two regions can be easily pulled apart to make the body of the object much wider. In contrast, the use of medial mesh as shown in \autoref{fig: deformation}(d) preserves the thickness very well in these regions.
For the fertility model, by fixing the base of the model and moving the handle on top of the model sideways, a nice bending effect shown in \autoref{fig: deformation}(d) can be obtained naturally with the medial-based method. On the other hand, the surface graph results in an undesirable global shear and more sophisticated handle manipulation is needed to achieve a similar bending effect.
It can be seen from both examples that the thickness of the object has been preserved well with the embedded volume graph (\autoref{fig: deformation}(c)), however, there is a lack of flexibility in bending and stretching perhaps because of the rigidity provided by the volume graph. The medial mesh offers such flexibility while maintaining the overall body thickness, as observed in most living objects.

%

As a final remark, Bloomenthal~\shortcite{Bloomenthal2002} considers the use of general medial representation for SSD and proposed a weighting scheme for mesh vertices by defining a convolution field around the medial axis. Yoshizawa et al.~\shortcite{Yoshizawa2007} achieve mesh deformation by deforming a skeleton mesh extracted from a triangle mesh, and reconstructing a surface defined by the deformed skeleton mesh using discrete differential coordinates. The skeleton mesh is a two-sided approximation of the medial axis which is geometrically a volume-collapsed closed mesh. Direct and coherent shape manipulation of this two-sided collapsed structure is nontrivial, and an additional stick-figure skeleton is required to first deform the skeleton mesh which in turn drives the surface deformation.

%


%

\section{Conclusion}

We have proposed the medial mesh as a new discrete approximation of the medial axis. The medial mesh defines a compact representation of a 3D shape as the union of simple enveloping primitives generated by swept spheres. We have also presented an efficient algorithm for computing a simplified and stable medial mesh of a given 3D shape. Experiments show that our method is efficient and robust. The medial meshes computed by our methods are much simpler and offer more accurate shape approximation than the results by previous methods. We have presented applications of the medial mesh to shape approximation and shape deformation. For shape approximation, the medial mesh is shown to provide a much better approximation than the existing methods using the union of spheres. For shape deformation, due to its simplicity and intrinsic nature the medial mesh demonstrates better performance in shape thickness preservation.

In summary, the medial mesh is a compact and stable representation of the medial axis, and thus has overcome the two notorious drawbacks of the medial axis, namely, instability and redundancy. Given the importance of the medial axis as a powerful intrinsic shape descriptor, we believe that the medial mesh will find more applications in shape modeling and analysis.

\paragraph{Future work and limitations}

The availability of the medial mesh as a simple and stable representation of the medial axis offers many research opportunities in shape modeling and analysis, such as shape recognition, shape matching, shape editing, shape segmentation, and collision detection. For collision detection, while the enveloping primitives induced by a medial mesh offer tighter bounding volumes, more research is needed to develop fast collision detection procedure for such primitives, since they are indeed more complex than bounding spheres which have been used extensively in collision detection.

Meanwhile, further improvement of the medial mesh is possible. The medial mesh encodes a 3D volume as the unions of its enveloping primitives. The boundary surface of this reconstructed volume is only $G^0$ continuous. So one future problem is to use piecewise smooth surface (such as subdivision surfaces) to approximate the medial axis to achieve higher order and smoother shape approximation. Another potential improvement of the medial mesh is its mesh connectivity. Our method for simplifying a medial mesh resembles the paradigm of mesh simplification based on edge merging. The resulting medial mesh, while stable and simple, may be further improved by optimizing its mesh connectivity or mesh vertex distribution, similar to the effect of surface remeshing.


\bibliographystyle{acmsiggraph}
\bibliography{mas}

\end{document}